\documentclass{CUP-JNL-EDS-preprint}

\usepackage{latexsym}
\usepackage{graphicx}
\usepackage{multicol,multirow}
\usepackage{amsmath,amssymb,amsfonts}
\usepackage{mathrsfs}
\usepackage{amsthm}
\usepackage{rotating}
\usepackage{appendix}
\usepackage[authoryear]{natbib}
\usepackage{ifpdf}
\usepackage[T1]{fontenc}
\usepackage{times}
\usepackage{newtxmath}
\usepackage{textcomp}%
\usepackage{xcolor}%
\usepackage{hyperref}
\usepackage{lipsum}

\usepackage{todonotes}
\usepackage{wasysym}
\usepackage{xspace}
\usepackage{cleveref}
\usepackage{ulem}

\articletype{APPLICATION PAPER}
\jname{Environmental Data Science}
\jyear{2023}
\jvol{4}
\jdoi{10.1017/eds.2020.xx}

\newcommand{\dO}{\ensuremath{\delta{}^{18}\textup{O}}\xspace}
\newcommand{\eighteenO}{\ensuremath{{}^{18}\textup{O}}\xspace}
\newcommand{\sixteenO}{\ensuremath{{}^{16}\textup{O}}\xspace}

\begin{document}

\begin{Frontmatter}

\title[Article Title]{Towards Learned Emulation of Interannual Water Isotopologue Variations in General Circulation Models}
\author*[1,2]{\mbox{Jonathan Wider}}\email{jonathan.wider@ufz.de}\orcid{0000-0002-5185-5768}
\author[1,2]{\mbox{Jakob Kruse}}
\author[1,3]{\mbox{Nils Weitzel}}\orcid{0000-0002-2735-2992}
\author[3]{\mbox{Janica C. Bühler}}\orcid{0000-0002-5100-4226}
\author[2]{\mbox{Ullrich Köthe}}\orcid{0000-0001-6036-1287}
\author[1,3,4]{\mbox{Kira Rehfeld}}\orcid{0000-0002-9442-5362}

\address[1]{\orgdiv{Institut für Umweltphysik}, \orgname{Ruprecht-Karls-Universität Heidelberg}, \orgaddress{\city{Heidelberg}, \country{Germany}}}
\address[2]{\orgdiv{Interdisziplinäres Zentrum für Wissenschaftliches Rechnen}, \orgname{Ruprecht-Karls-Universität Heidelberg}, \orgaddress{\city{Heidelberg}, \country{Germany}}}
\address[3]{\orgdiv{Department of Geosciences}, \orgname{University of Tübingen}, \orgaddress{\city{Tübingen}}, \country{Germany}}
\address[4]{\orgdiv{Department of Physics}, \orgname{University of Tübingen}, \orgaddress{\city{Tübingen}}, \country{Germany}}

\received{30 January 2023}
\revised{--}
\accepted{--}
\authormark{Wider et al.}

\abstract{
Simulating abundances of stable water isotopologues, i.e. molecules differing in their isotopic composition, within climate models allows for comparisons with proxy data and, thus, for testing hypotheses about past climate and validating climate models under varying climatic conditions.
However, many models are run without explicitly simulating water isotopologues.
We investigate the possibility to replace the explicit physics-based simulation of oxygen isotopic composition in precipitation using machine learning methods. These methods estimate isotopic composition at each time step for given fields of surface temperature and precipitation amount.
We implement convolutional neural networks (CNNs) based on the successful UNet architecture and test whether a spherical network architecture outperforms the naive approach of treating Earth's latitude-longitude grid as a flat image.
Conducting a case study on a last millennium run with the iHadCM3 climate model, we find that roughly 40\% of the temporal variance in the isotopic composition is explained by the emulations on interannual and monthly timescale, with spatially varying emulation quality. 
A modified version of the standard UNet architecture for flat images yields results that are equally good as the predictions by the spherical CNN.
We test generalization to last millennium runs of other climate models and find that while the tested deep learning methods yield the best results on iHadCM3 data, the performance drops when predicting on other models and is comparable to simple pixel-wise linear regression.
An extended choice of predictor variables and improving the robustness of learned climate--oxygen isotope relationships should be explored in future work.
}

\policy{
Information on the hydrological cycle is imprinted onto the isotopic composition of precipitation, which subsequently is oftentimes preserved in natural climate archives like speleothems or ice deposits. Some climate models, so-called isotope-enabled general circulation models (iGCMs), simulate isotopes explicitly and, thus, allow comparing climate model output under paleoclimate scenarios to samples taken from natural climate archives. However, isotopes are not included in most climate simulations due to computational constraints or the complexity of their implementation. We test the possibility of using machine learning methods to infer the isotopic composition from surface temperature and precipitation amounts, which are standard outputs for a wide range of climate models.
}

\end{Frontmatter}

\section[Introduction]{Introduction}
\label{sec:intro}

Reliable analysis of current climate change, as well as robust prediction of future Earth system behavior, have become a crucial foundation for all endeavors to protect humanity's prosperity, mitigate ecological disasters, or formulate plans for adaptation \citep{langsdorfClimateChange20222022}.
This analysis hinges on an accurate understanding and modeling of complex mechanisms in the climate system, which in turn relies on knowledge of the system's past behavior.
To analyze past climatic conditions outside the comparatively short period of instrumental measurements, we depend on environmental processes recording and preserving information on the climate system in natural ``climate archives''.
One way to recover past climate information from such archives is to measure the relative abundance of isotopes, particularly of the isotopes of the constituents of water molecules \citep{mookEnvironmentalIsotopesHydrological2000}.
Due to differences in mass, molecules with varying isotopic compositions, so-called isotopologues, differ in their behavior in chemical reactions and phase transitions. For the special case of water, molecules containing heavy \eighteenO atoms, further denoted heavy isotopes, evaporate slower but condensate faster than ones containing the lighter \sixteenO.
These effects are imprinted on the global hydrological cycle. The resulting patterns of the isotopic composition of precipitation depend on many variables such as precipitation amount, temperature, relative humidity, and the circulation of the atmosphere \citep{dansgaardStableIsotopesPrecipitation1964}.
This makes heavy isotopes in water an important tracer of the hydrological cycle and consequently a valuable proxy for past climatic changes.

Isotopic abundances are canonically expressed in the delta notation.
For stable oxygen isotopes \eighteenO and \sixteenO, this is given by
\begin{equation}
    \label{eq:d18O}
    \dO = \left( \tfrac{[\eighteenO_\text{sample}]}{[\sixteenO_\text{sample}]} \middle/ \tfrac{[\eighteenO_\text{reference}]}{[\sixteenO_\text{reference}]} \right) - 1 [\permil].
\end{equation}
Here the ratio of concentrations of the isotopic species in a given sample is compared to a defined reference standard.
For \dO of precipitation, this standard is an artificially created sample with an isotopic composition that is typical for ocean surface water \citep{baertschiAbsolute18OContentStandard1976}.

One important task in paleoclimatology is to test whether hypotheses about the past climate are compatible with proxy data like \dO measured in natural climate archives \citep[e.g.][]{buhlerInvestigatingStableOxygen2022}.
To compare simulations of hypothetical climate states to those measurements, a special sub-type of climate models, so-called isotope-enabled General Circulation Models (iGCMs), was developed.
They explicitly simulate isotopic compositions by following the isotopic water species through the hydrological cycle \citep{bradyConnectedIsotopicWater2019, tindallStableWaterIsotopes2009, yoshimuraHistoricalIsotopeSimulation2008, coloseInfluenceVolcanicEruptions2016, wernerGlacialInterglacialChanges2016b}.
However, many climate models and climate model simulations exist that do not include information on water isotopologues.
Simulating \dO is costly because it typically requires duplicating large parts of the water cycle for each simulated water species \citep{tindallStableWaterIsotopes2009}.
In light of recent advances in data science, the question arises whether this isotopic output can instead be emulated using machine learning (ML) models that infer the \dO at each location from other climate variables after a model run is finished. We thus call this approach ``offline-emulation''. Conducting the emulation ``offline'', i.e. not coupled to the climate simulation, is possible because isotopes are passive tracers of the hydrological cycle that have no feedback onto the climate system.

Within this study, we narrow the broad task of ``offline-emulation'' by making a number of choices for the learned isotope emulation.
The first choice is to only emulate the isotopic composition of precipitation.
This is the quantity that is also output by iGCMs; it neglects all processes that might disturb the signal until it is stored in a climate archive \citep[see e.g. ][]{casadoArchivalProcessesWater2018}.
Because observations of isotopes in precipitation are sparse and only exist starting from the 1960s \citep{IAEAWMO2020}, we conduct our experiments entirely on simulated data. We limit ourselves to using surface temperature and precipitation amount as the two fundamental predictor variables since these variables possess strong correlations to $\dO$ that are well known experimentally \citep{dansgaardStableIsotopesPrecipitation1964} and from simulations (see \Cref{fig:iHadCM3_stats}, \textsf{C}) and are frequently simulated in climate models.

We decide to emulate yearly \dO data from last millennium (850 CE to 1849 CE) climate simulations. This is motivated by the combination of the high data availability of simulation runs of sufficient length, and the archiving resolution of paleoclimate records during this time period which is typically between monthly and sub-decadal. We also contrast the yearly emulation results with experiments using monthly resolution.

As a measure of emulator performance, we will use the $R^2$ score, which measures the fraction of explained temporal variance, as detailed in \Cref{sec:metric}.
While we use ML methods that exploit spatial correlations in the data by design, we leave explicit incorporation of temporal correlations largely to future investigation.

\vspace{1em}
Working within these constraints, our paper presents the following contributions:
\begin{itemize}
    \item We train a deep neural network to estimate stable oxygen isotopes in precipitation (\dO) given surface temperature and precipitation and compare to common regression baselines.
    \item To respect the underlying geometry of the climate model data, we investigate the performance of a spherical network architecture.
    \item We present cross-model results, where a regressor trained on simulated data from one climate model is used to emulate \dO in a run from a different model.
\end{itemize}

\section[Data and Methodology]{Data and Methodology}
\begin{figure}[t]%
    \FIG{\includegraphics[width=\textwidth]{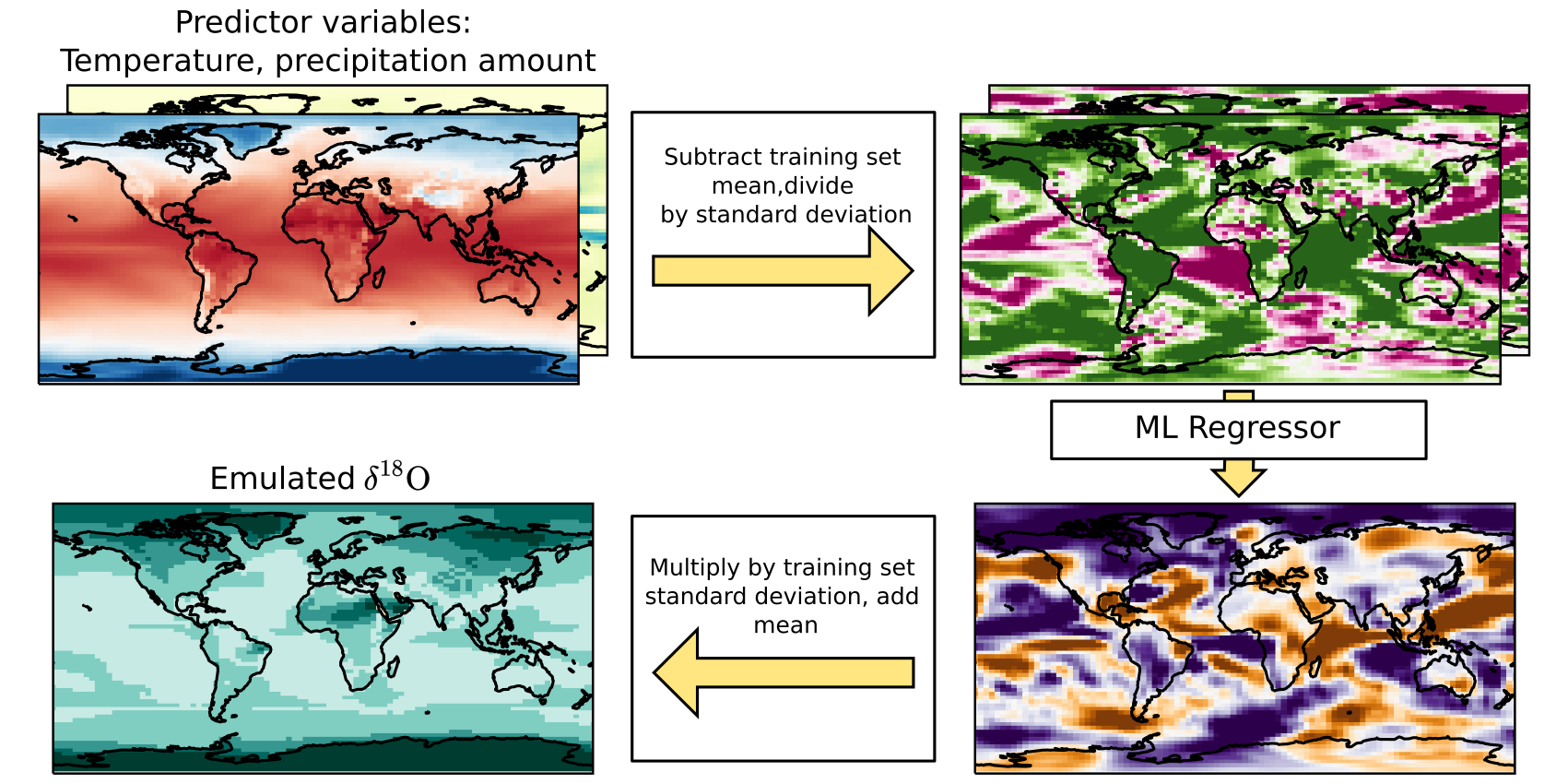}}
    {\caption{Our approach to the emulation of \dO in precipitation:
    for each time step, we use surface temperature and precipitation amount as predictor variables. Subsequently, the data is standardized pixel-wise by subtracting the mean and dividing it by the standard deviation at each pixel (top right). Means and standard deviations are based on the training set of the investigated climate model simulation. We use a machine learning emulation model (ML Regressor) to obtain a standardized estimate for \dO. The emulator output (bottom right) is then de-standardized using the training set mean and standard deviation of \dO at every pixel, to arrive at the final emulation result (bottom left).
    When applying the ML model to data from climate models other than the one that was used for training (e.g. in the cross-comparison experiment in \Cref{sec:results-cross-comparison}) we use the mean and standard deviation from the training set of the new model}
    \label{fig:overview}}
\end{figure}

    Our approach to emulating \dO is sketched in \Cref{fig:overview}. For each time step, we start with variables that we know are statistically related to \dO, namely surface temperature and precipitation amount. All variables are standardized pixel-wise, i.e. we subtract the mean and divide by the standard deviation, both computed from the training set. We then estimate the standardized spatial field of \dO from the predictor variables by training a machine learning (ML) regression model. Subsequently, the standardization for the inferred \dO is reversed, resulting in our estimate for the isotopic composition.
    
    \subsection{Data}
\label{sec:data}

We use data from the isotope-enabled version of the Hadley Center Climate Model version 3 climate model \citep[hereafter iHadCM3,][]{tindallStableWaterIsotopes2009}. iHadCM3 is a fully coupled atmosphere-ocean general circulation model (AOGCM). The horizontal resolution of iHadCM3 is 3.75° in the longitudinal direction, and 2.5° in the latitudinal direction. We exclude -90° and 90° from the latitudinal values because \dO is not simulated at these latitudes. We focus on the last millennium (850 CE to 1849 CE), which is characterized by a stable climate with variability on interannual-to-centennial timescales, but no major trends \citep{jungclausPMIP4ContributionCMIP62017}. Additionally, the last millennium is well documented in climate archives and observations \citep{PAGES2k2019, Konecky2020, Comas-Bru2020, HadCRUT4}.

\begin{figure}[t]%
    \FIG{\includegraphics[width=.9\textwidth]{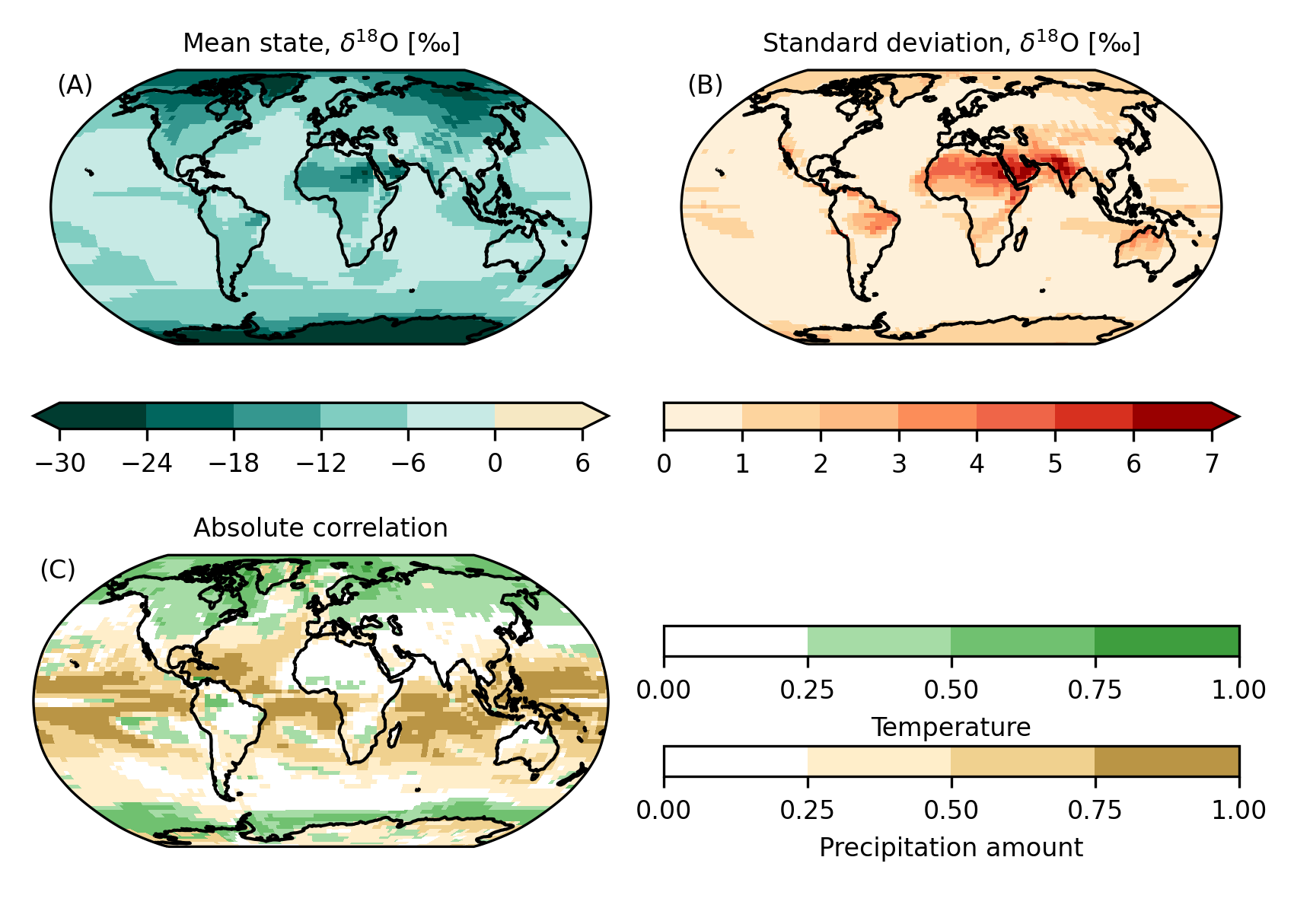}}
    {\caption{
        Statistical properties of the iHadCM3 \dO data:
        \textsf{\textup{(A)}} mean state of isotopic composition (\dO) in the precipitation in precipitation and
        \textsf{\textup{(B)}} standard deviation of \dO on an annual timescale.
        \textsf{\textup{(C)}} absolute correlations of \dO with temperature (green) and precipitation amount (brown) on interannual timescale; for each grid cell only the stronger of the two is shown%
    }
    \label{fig:iHadCM3_stats}}
\end{figure}

Diagnostics of the iHadCM3 data set are visualized in \Cref{fig:iHadCM3_stats}. As can be seen from \Cref{fig:iHadCM3_stats} \textsf{B}, the standard deviation of the simulated \dO is large over dry regions like the Sahara desert or the Arabian peninsula. This is partly related to the way \dO is computed in the climate models: in these regions the abundances of \eighteenO and \sixteenO are both small because of generally low precipitation amounts, leading to numerically unstable ratios and missing values on the monthly time scale. Overall, 0.3\% of the \dO values are missing on the monthly timescale, with a strong clustering in the regions with numerical instabilities described above (compare \Cref{fig:missing_pixels}). We take this into account by adapting the loss we use to train our ML methods to deal with missing values, as described in \Cref{sec:loss}.

To test the extrapolation and robustness of our emulator, we use last millennium simulations of three other climate models: AGCM Scripps Experimental Climate Prediction Center’s Global Spectral Model \citep[hereafter isoGSM,][]{yoshimuraHistoricalIsotopeSimulation2008}, iCESM1 version 1.2 \citep[hereafter iCESM,][]{bradyConnectedIsotopicWater2019}, and ECHAM5/MPI-OM \citep[hereafter ECHAM5-wiso,][]{wernerGlacialInterglacialChanges2016b}. While iCESM and ECHAM5-wiso are fully coupled AOGCMs, isoGSM is an atmospheric GCM forced by sea surface temperatures and sea ice distributions of a last millennium run with the CCSM4 climate model \citep{landrumLastMillenniumClimate2013}. We re-grid the other climate model simulations to the iHadCM3 grid using bilinear interpolation from the CDO tool set \citep{schulzweidauweCDOUserGuide2020}.

All data sets are freely available at \url{https://doi.org/10.5281/zenodo.7516327} and described in detail in \citet{buhlerInvestigatingStableOxygen2022}.\footnote{\citet{buhlerInvestigatingStableOxygen2022} also investigate a fifth climate model, GISS ModelE2-R \citep{coloseInfluenceVolcanicEruptions2016}, which we excluded from our study because of physically implausible trends in polar regions in the corresponding model run.}

\subsubsection{Pre-processing}
We apply the following pre-processing steps to the climate simulation data:

\begin{itemize}\vspace{-.5em}
    \item We set valid ranges for all variables, thereby excluding implausibly large or small values, using the following choices: surface temperature range: $[173, 373]$ K, \dO range: $[-100,100] \permil$, precipitation amount: $[-1,10000] \frac{\text{mm}}{\text{month}}$. Wide ranges are chosen because we aim to exclude only implausible values that might deteriorate emulator performance without artificially removing model deficiencies. Thus, we also keep small negative precipitation values that climate models might produce due to numerical inaccuracies in rare occasions.
    \item Time steps with missing values in the predictor variables are excluded from the data set. This leads to the exclusion of 31 of the 12000 monthly time steps of iHadCM3.
    \item We form yearly averages from monthly data. Missing \dO data points are omitted in the yearly averaging. We argue that this does not impact our results negatively, because the invalid 0.3\% of \dO values cluster in regions, where due to numerical instabilities in the ``ground truth'' iHadCM3 simulation, learning a physically consistent emulation would not have been possible anyway (compare \Cref{fig:missing_pixels}).
    \item We re-grid the yearly data sets to the irregular grid on which the investigated spherical network operates (see \Cref{sec:method}) using a first-order conservative remapping scheme \citep{schulzweidauweCDOUserGuide2020}.    
    \item We split the data into test and training sets. We use 850-1750 CE for training and 1751-1849 CE for testing. The data are split chronologically instead of randomly to make the test and training set as independent as possible, and prevent the network from exploiting auto-correlations from previous or subsequent time steps. If a validation set (used for making choices of ML hyperparameters) is needed, we split off 10\% of the training set randomly unless specified otherwise.
    \item Before the ML methods are applied, the data are standardized pixel-wise by subtracting the training set mean and dividing by the standard deviation of the corresponding climate model, as visualized in \Cref{fig:overview}.
\end{itemize}

    \subsection{Methodology}
\label{sec:method}

To obtain a spatially consistent emulation, and to utilize the fact that the local statistical relations between \dO and the predictor variables are similar in many grid boxes on Earth's surface, we choose two approaches based on Convolutional Neural Networks (CNNs). Both utilize the successful UNet architecture \citep{ronnebergerUNetConvolutionalNetworks2015}, whose multi-scale analysis can simultaneously capture fine structure variations and utilize large-scale contextual information. UNet architectures have been successfully applied in a climate science context before \citep[e.g.][]{kadowArtificialIntelligenceReconstructs2020}. The first of our two approaches treats data on the latitude-longitude grid as a flat image. The second explicitly incorporates the spherical geometry of the data.

\subsubsection{Flat network}
\label{sssec:flat_network}
Because our data naturally lie on the surface of a sphere, distortions arise when treating the equally spaced longitude-latitude grid as a flat image using e.g.~a plate carrée projection (lat/lon projection). We test if we can still obtain reasonable results with this naive setup. Furthermore, we try to partially remedy the effects of the distortions within the ``flat'' approach, by modifying the standard UNet architecture in three ways :

\begin{itemize}
    \item We use area-weighted loss functions.
    \item We use periodic padding in the longitudinal direction, i.e. we append the rightmost column to the very left of the plate carrée map (and vice versa) before computing convolutions. Thereby we assure continuity along the 0°-360° coordinate discontinuity.
    \item We incorporate CoordConv \citep{liuIntriguingFailingConvolutional2018}, a tweak to convolutional layers that appends the coordinates to the features input into each convolution, thus, allowing networks to learn to break translational symmetry if necessary.
\end{itemize}

\subsubsection{Spherical network}
As a more sophisticated technique, a multitude of approaches to directly incorporate the spherical nature of data into a neural network architecture has been proposed \citep{cohenGaugeEquivariantConvolutional2019, cohenSphericalCNNs2018, coorsSpherenetLearningSpherical2018, defferrardDeepSphereGraphbasedSpherical2020, estevesLearningEquivariantRepresentations2018, lamGraphCastLearningSkillful2022}. We reproduce the approach of \citet{cohenGaugeEquivariantConvolutional2019}, where the network operates on an icosahedral grid, with grid boxes centered on the vertices. Using the icosahedron offers a straightforward way to increase or decrease resolution for a UNet-like design, as we can recursively subdivide each of its triangles into four smaller triangles, projecting all newly created vertices onto the sphere again. We denote the number of recursive refinements of the grid as $r$, with $r=0$ identifying the grid containing only the twelve vertices of the regular icosahedron. As the refined icosahedral grid is locally very similar to a flat hexagonal grid, we can use an appropriately adapted implementation of the usual efficient way to compute convolutions. Additionally, the architecture of \citet{cohenGaugeEquivariantConvolutional2019} is equivariant to a group of symmetry transformations, meaning that if the input to the CNN is transformed by an element of the symmetry group, the output transforms accordingly. This fits well with the approximate symmetries present in the Earth system, like symmetry to reflections on the equatorial plane or rotations around the polar axis. We validate our implementation of the method on a toy problem  described by \citet{cohenGaugeEquivariantConvolutional2019}: the classification of handwritten digits projected onto a spherical surface. We obtain results that are comparable to those reported by \citet{cohenGaugeEquivariantConvolutional2019}, see \Cref{sssec:validation} for more details. 

\subsubsection{Loss function}
\label{sec:loss}
To train our UNet architectures for isotope emulation, we use a weighted mean squared error loss between the standardized \dO ground truth $Y$ and the predicted values $\hat{Y}$:

\begin{equation}
    \label{eq:weighted_masked_mseloss}
    L(Y, \hat{Y}) = \frac{1}{b}\sum_{i=1}^{b} \frac{1}{|\mathcal{G}_i|} \sum_{j \in \mathcal{G}_i}^{|\mathcal{G}_i|} w_j \Bigl( Y_{i,j} - \hat{Y}_{i,j} \Bigr)^2 ,
\end{equation}
where the loss is averaged over a batch of size $b$ and the set of valid grid boxes $\mathcal{G}_i$ at time step~$i$. A grid box is valid if the simulated ground truth data has no missing value at this time step in this grid box. $|\mathcal{G}_i|$ denotes the cardinality of $\mathcal{G}_i$ and $w_j$ are weighting coefficients. For the convolutional UNet working on the plate carrée projection, we choose $w_j$ to be proportional to the cosine of the latitude of the center of grid cell~$j$, which is an approximation of the physical size of the grid cell. We rescale the weights, such that they sum to the total number of grid boxes. For the icosahedral UNet, all grid boxes are of approximately equal size. Therefore, no weighting is applied and $w_j$ is a constant independent of $j$.

\subsubsection{Baselines}
In addition to the UNet models, we implement three simple baseline models to assess the relative benefit of complex and deep models in our emulation problem. These baselines are:
\begin{itemize}

    \item \textbf{Grid-box-wise linear regression}, the simplest conceivable model:
    regress \dO on temperature and precipitation amount in a separate model for each grid box.

    \item \textbf{Grid-box-wise random forest regression model}:
    in contrast to the linear regression baseline, we train a single random forest \citep{breimanRandomForests2001} to make predictions on all grid boxes. To allow the model to learn spatially varying relationships, we include the coordinates as predictor variables.\footnote{We encode each longitude $\phi$ as two values $\sin(\phi)$ and $\cos(\phi)$ to avoid the discontinuity at 0\textdegree/360\textdegree.}

    \item \textbf{Grid-to-grid approach (PCA regression)}:
    relations between \dO and other climatic variables tend to behave similarly over large areas (see \Cref{fig:iHadCM3_stats} \textsf{C}),
    justifying a dimension reduction of the input and output spaces before applying a multivariate linear regression. This is implemented by computing the principal components of the input and output spaces. Schematically, the computation goes as follows: $X \overset{\mathrm{PCA}_X}{\mapsto} C_X \overset{\text{lin.~reg.}}{\mapsto} \hat{C}_Y \overset{\mathrm{PCA}_Y^{-1}}{\mapsto} \hat{Y}$. Approximately optimal numbers of principal components are obtained as follows: we iterate over a $50 \times 50$ logarithmically spaced grid of candidate values for the number of input and output principal components. For each configuration, the emulation model is trained and its performance is measured on a held-out validation set. We then select the combination of numbers of input and output principal components which yields the best results on the validation set. As a last step, the selected model is retrained, now including the validation set data. Principal component analysis can be performed on arbitrary grids, which makes it equally applicable to the projected 2D data and the icosahedral representation.

\end{itemize}

\subsubsection{Metrics}
\label{sec:metric}
The metric we use for evaluating emulation approaches is the $R^2$ score, also called the ``coefficient of determination'', which quantifies what fraction of the temporal variance in the test set is explained by the ML estimate in each grid box.
The $R^2$ score compares the \dO ground truth $Y_j$ and an estimate $\hat{Y}_j$ in a given grid box $j$ as
\begin{equation}
    \label{eq:r2_score}
    R^2(Y_j, \hat{Y}_j) = 1 - \frac{\text{MSE}(Y_j, \hat{Y}_j)}{\sigma_j^2} ,
\end{equation}
where $\text{MSE}(Y_j, \hat{Y}_j)$ is the mean squared error and $\sigma_j^2$ the variance of the test set ground truth, both taken over the time axis at grid box~$j$. A value of $R^2 = 1$ indicates perfect emulation, while a model that simply outputs the temporal mean at every time step has $R^2 = 0$. The score can become arbitrarily negative.

Additionally, we compute the Pearson correlation coefficient between the true and emulated time series at some grid boxes. To select time steps in which a method's performance is particularly strong or weak, we calculate the Anomaly Correlation Coefficient (ACC) between emulation and ground truth. ACC is defined as the Pearson correlation coefficient between the true and emulated anomaly patterns for a given time step. Anomalies are computed with respect to the training set mean. 

If error intervals on performance metrics are given, unless stated otherwise, they are $1\sigma$ intervals computed over a set of ten runs. Thus, the uncertainties only account for the uncertainty of the stochastic aspects of the ML model parameter optimization, discarding any uncertainty that is related to the data.

Implementation details for training and configuration of the ML methods are provided in \Cref{ssec:implementation_details} and code to reproduce our experiment is freely available at \url{https://github.com/jonathanwider/isoEm}

\section[Results]{Results}
\label{sec:results}
We structure the Results section as follows. First, we give a detailed spatiotemporal overview to illustrate the characteristics of the ML-based emulation results. To this purpose, we use the best performing emulation method as an example. Subsequently, we compare emulation methods amongst each other, contrasting deep architectures and baselines as well as ``flat'' and ``spherical'' approaches. We follow up with a range of sensitivity experiments, and conclude by conducting a cross-model experiment, i.e. we train a ML model on data from one climate model and then use the trained model to emulate \dO in other climate model simulations.

\subsection{Spatiotemporal Overview of Emulation Results}
\label{sec:results-main}
In \Cref{sec:results-ablations}, we will discover that the  best performing ML emulation method, a deeper version of the flat UNet architecture, reaches an average $R^2$ score of $0.389 \pm 0.006$ on the plate carrée grid. This means that in the global average almost $40\%$ of the temporal variance in the test set is explained by our emulation on the interannual timescale. We use this best ML method to introduce spatial and temporal characteristics of the emulation. 

\begin{figure}[t]%
\FIG{\includegraphics[width=\textwidth]{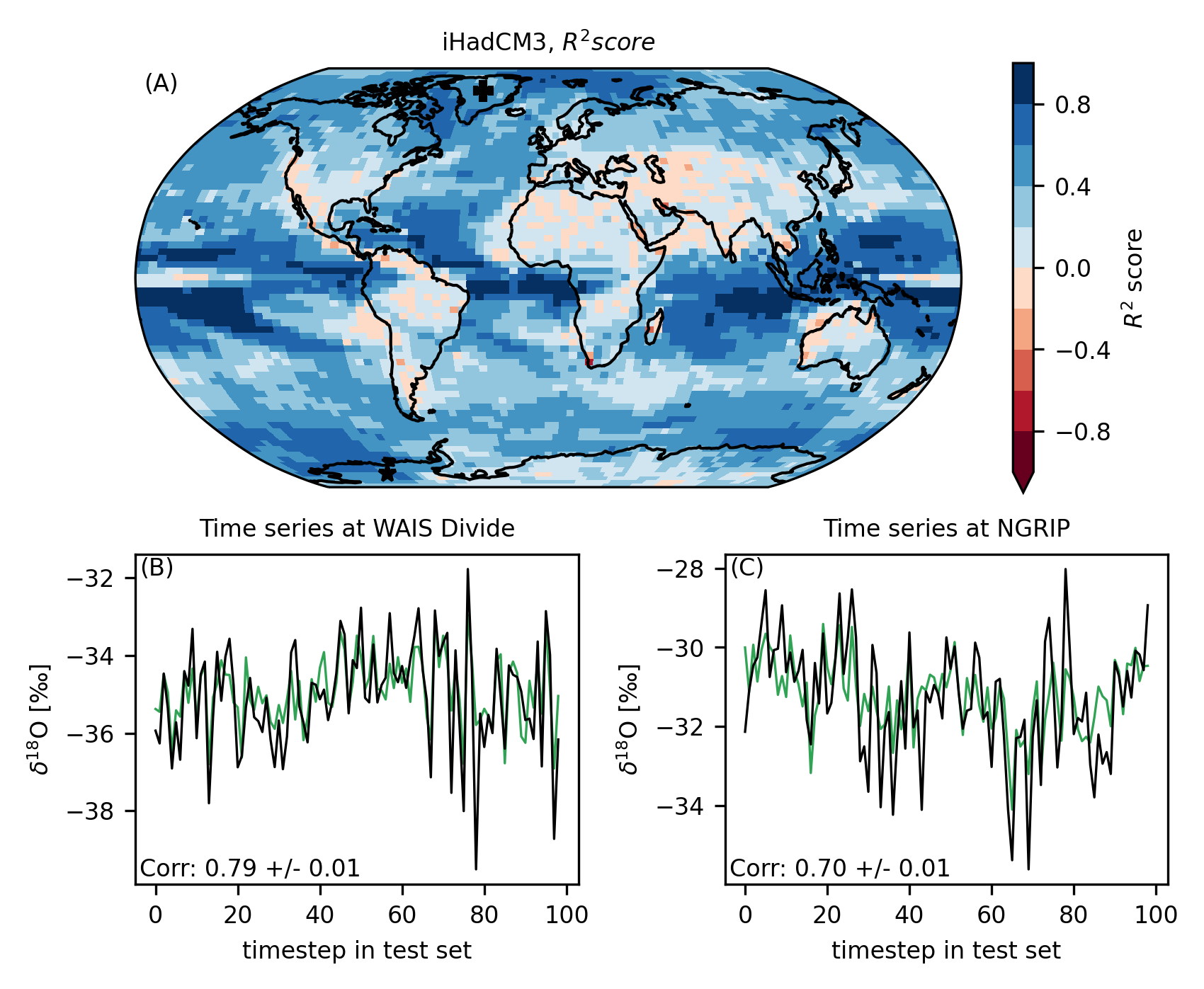}}
{\caption{Test set emulation performance of the best ML emulation method. The bluer the colors, the better the emulation. Blue colors indicate regions in which the performance is better than a trivial baseline model that returns the correct test set mean at every time step. This plot displays the average of the $R^2$ scores over ten runs. Additionally, we show the time series of the ML emulation (green, mean over ten runs) and the true simulation data (black) for grid boxes next to two ice core drilling sites. Panel \textsf{\textup{(B)}} "NGRIP" (Greenland). Panel \textsf{\textup{(C)}} "WAIS Divide" (West Antarctica)}
\label{fig:iHadCM3_main}}
\end{figure}

The prediction quality varies spatially, as shown in \Cref{fig:iHadCM3_main}\textsf{\textup{A}}. $R^2$ scores of $0.6$ or larger are reached in $18.5\%$ of the grid cells, and $R^2 \leq 0$ for only $5.4\%$ of grid cells. The best results are achieved over tropical oceans, which are regions with strong correlations of \dO and precipitation amounts. Performance is good over large parts of the Arctic and over western Antarctica as well, which is important because these regions are especially relevant for the comparison with \dO measurements from ice cores. We illustrate the performance in these regions by comparing emulated and ground truth time series in the grid boxes closest to two ice core drilling sites in panels \textsf{\textup{B}} and \textsf{\textup{C}} of \Cref{fig:iHadCM3_main}: the North Greenland Ice Core Project 
\citep[``NGRIP'', 75.1° N, 42.3° W, ][]{berggren600yearAnnual10Be2009} and the West Antarctic Ice Sheet Divide ice core project \citep[``WAIS Divide'', 79.5° S, 112.1° W, ][]{buizertWAISDivideDeep2015}. For these drilling sites, the correlation between our emulation and the exact output time series of an isotope-enabled climate model exceeds $70\%$.

In general, spatial variations in performance follow the correlation structure between \dO and the predictor variables (\Cref{fig:iHadCM3_stats}\textsf{C}): in regions with strong absolute correlations between \dO and surface temperature or precipitation amount, the $R^2$ scores are higher than in regions where none of the predictor variables is strongly correlated with \dO. Thus, performance is worse over landmasses, especially in the low and mid-latitudes.

\begin{figure}[t]%
\FIG{\includegraphics[width=\textwidth]{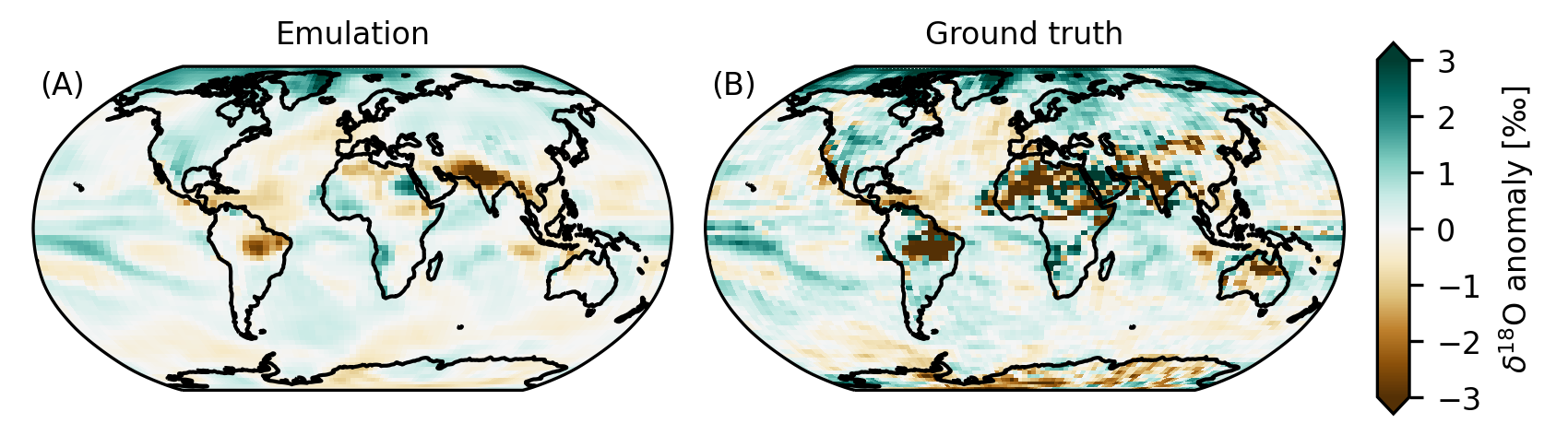}}
{\caption{Typical emulation results on iHadCM3 data set: We show anomalies as they are output by the ML-emulator (``Emulation'') and the ``true'' result in the simulation data set (``Ground truth''). The anomalies are computed with respect to the training set mean. For the selected time step, the anomaly correlation coefficient (ACC) reaches its median value}
\label{fig:iHadCM3_anomalies}}
\end{figure}

Next, we visualize emulation and climate model output for individual time steps. For a year with typical emulator performance\footnote{We chose the median in terms of anomaly correlation coefficient (ACC).}, we plot emulated (panel \textsf{\textup{A}}) and simulated (panel \textsf{\textup{B}}) anomalies in \Cref{fig:iHadCM3_anomalies}. We can see that the large-scale patterns match well between emulation and simulation: there are strong positive anomalies over the Arctic, related to positive temperature anomalies in this time step, and the large-scale structure over the Pacific is captured as well. Strong negative anomalies over parts of South America and northern India and Pakistan are reproduced. 
Emulation and ground truth simulation differ in their fine-scale structure: the ground truth is generally less smooth than the emulation and seems particularly noisy over some dry regions like the Sahara and the Arabic desert. In these regions, there is a potential for numerical inaccuracies in the isotopic component of climate models, due to small abundances of each isotopic species, and it is hard to untangle which parts of the ``noisy'' signal have a climatic origin and which parts are simulation artifacts. A part of the overall smoother nature of the UNet regression results can be attributed to the MSE Loss giving a large (quadratic) penalty for strong deviations from the true values, thus, priming the network against predicting values in the tails of the distribution. \Cref{fig:iHadCM3_anomalies_all} compares emulation and simulation for three additional time steps: time steps in which the emulation works particularly well or poorly, and a climatically interesting year: 1816~CE, the ``year without a summer'' \citep{luterbacherYearSummer2015}, which is caused by a volcanic eruption included in the volcanic forcing of the iHadCM3 simulation. For 1816~CE, we observe that the emulator reproduces a strong negative \dO anomaly in regions where \dO is primarily influenced by temperature, namely in the Arctic, Northern North America and Siberia.

\subsection{Comparing Machine Learning Methods}
\label{sec:results-ml-methods}

The ML emulation models (UNet architectures and simpler baselines) differ in the quality of their emulation. In the following, we compare the methods amongst each other. For details on the training procedures, network architectures, and method implementations, see \Cref{ssec:implementation_details}. We also address the question of whether using an inherently spherical approach is beneficial over treating the latitude-longitude grid as ``flat''. However, the comparison is not trivial: the approaches are developed for data on different grids (plate carrée and icosahedral) and the necessary interpolations may deteriorate performance. Thus, we compute performances on both grids, interpolating the predictions from one grid to another. Results for the globally averaged $R^2$ scores are shown in \Cref{tab:architectures}. 
The best model in the comparison is the ``modified'' version of the flat UNet that includes the three modifications (area-weighted loss, adapted padding, CoordConv) described in \Cref{sec:method}. The effect of the individual modifications is detailed in \Cref{tab:modifications} and \Cref{fig:modifications_flat_unet}.

All UNet architectures outperform all baseline architectures that operate on the same grid. The best UNet method explains $7\%$ more of the test set variance than the best baseline model, PCA regression. The other baseline models perform worse. In particular, it seems that the random forest baseline, which regresses on a pixel-to-pixel level is not able to capture the spatially varying relationships between \dO and the predictor variables surface temperature and precipitation amount well enough, even when including coordinates as additional inputs. The spatial performance differences between the UNet methods and the best baselines are visualized in \Cref{fig:differences_to_baselines}. The improvements by the UNets are largest over oceans.

On the icosahedral grid, the icosahedral UNet and the modified flat UNet achieve $R^2$ scores that are not significantly different. On the plate carrée grid, however, the results of the icosahedral UNet are much worse. This drop can largely be attributed to the interpolation method (see \Cref{fig:results_flat_grid_interpolation}): on the plate carrée grid, neither training data nor results of the flat UNet are interpolated, while for the icosahedral UNet interpolations are necessary in both cases.

\begin{table}[t]
\tabcolsep=0pt%
\TBL{\caption{Globally averaged $R^2$ scores for the different ML-emulation methods. Results are calculated for the icosahedral grid that the method of \cite{cohenGaugeEquivariantConvolutional2019} operates on and the plate carrée grid. When a method works with data on the other grid, the emulated data is interpolated. "Flat UNet, unmodified" and "Flat UNet, modified" refer to the flat network architecture described in \Cref{sssec:flat_network}, either not applying or applying the modifications to remedy projection artifacts described in that chapter.}
\label{tab:architectures}
}
{\begin{fntable}
\begin{tabular*}{\textwidth}{@{\extracolsep{\fill}}lcc@{}}\toprule%
\TCH{Emulation Method} & \TCH{$R^2$score, plate carrée grid} & \TCH{$R^2$score, icosahedral grid}\\
\midrule
\TCH{Flat UNet, unmodified} & $0.352 \pm 0.015$ & $0.374 \pm 0.017$\\
\TCH{Flat UNet, modified} & $\textbf{0.377} \pm 0.005$ & $\textbf{0.402} \pm 0.006$\\
\TCH{Flat random forest baseline} & $0.212$ & $0.256$\\
\TCH{Flat linear regression baseline} & $0.251$ & $0.274$\\
\TCH{Flat PCA regression baseline} & $0.303$ & $0.332$\\
\TCH{Icosahedral UNet} & $0.126 \pm 0.011$ & $0.396 \pm 0.009$\\
\TCH{Icosahedral PCA regression baseline} & $0.076$ & $0.339$\\
\botrule
\end{tabular*}%
\end{fntable}}
\end{table}

\subsection{Sensitivity experiments}
\label{sec:results-ablations}
We conduct a range of sensitivity experiments, to test a) the influence of each predictor variable on the results, b) whether we can further improve the performance of our ML method and c) whether emulation quality varies with timescale.

First, we use the modified flat UNet architecture as employed in \Cref{sec:results-ml-methods} and test, how the results differ if we exclude one of the predictor variables. The globally averaged $R^2$ score on the plate carrée grid drops from $0.377 \pm 0.005$ if both precipitation and temperature are used to $0.327 \pm 0.006$ when using only precipitation and to $0.251 \pm 0.004$ when only using temperature. The spatial differences in emulation quality follow the large-scale behavior of the correlation structure in panel \textsf{\textup{C}} of \Cref{fig:iHadCM3_stats}. When precipitation is excluded, the performance decreases most over low latitudes, while the $R^2$ score drops over polar regions without temperature. This is visualized in \Cref{fig:differences_variables}.

To potentially improve the emulation results even further, we create variations of the modified flat UNet architecture: a ``wider'' version in which the number of computed features per network layer is doubled ($R^2 = 0.386 \pm 0.008$, plate carrée grid), and a ``deeper'' version with six additional network layers\footnote{I.e.~one additional ``depth step'' in \Cref{fig:unet_flat}}, which obtains $R^2 = 0.389 \pm 0.006$ (plate carrée grid), both improving over the default choice by roughly $0.01$ . Additionally, we test, whether results could be improved by tuning the learning rate of the employed optimizer by testing a grid of 20 logarithmically spaced values between $10^{-4}$ and $10^{-1}$. The performance is best for learning rates between $10^{-3}$ and $10^{-2}$. However, no substantial improvements over the default parameter choice were reached in the limited range of tested values.

The monthly timescale differs from the interannual scale by a pronounced seasonal cycle of \dO in many regions. Thus, even a simple climatology can explain a part of the variability in \dO. To exclude this trivially explainable part from the computation of the $R^2$ score, we compute the score separately for each month. Results are similar to the results on the interannual scale with roughly $40\%$ of variance explained. 
The higher time resolution suggests exploring whether the emulation can profit from taking the temporal context into account. We test this by including not only the temperature and precipitation of the current time step but also of the previous month as inputs to the emulation of \dO. Results do not improve strongly, however, possibly because the investigated timescale is still larger than the average atmospheric moisture residence time \citep{trenberthAtmosphericMoistureResidence1998}.

\subsection{Cross-Model Comparison}
\label{sec:results-cross-comparison}
\begin{figure}[t]%
\FIG{\includegraphics[width=\textwidth]{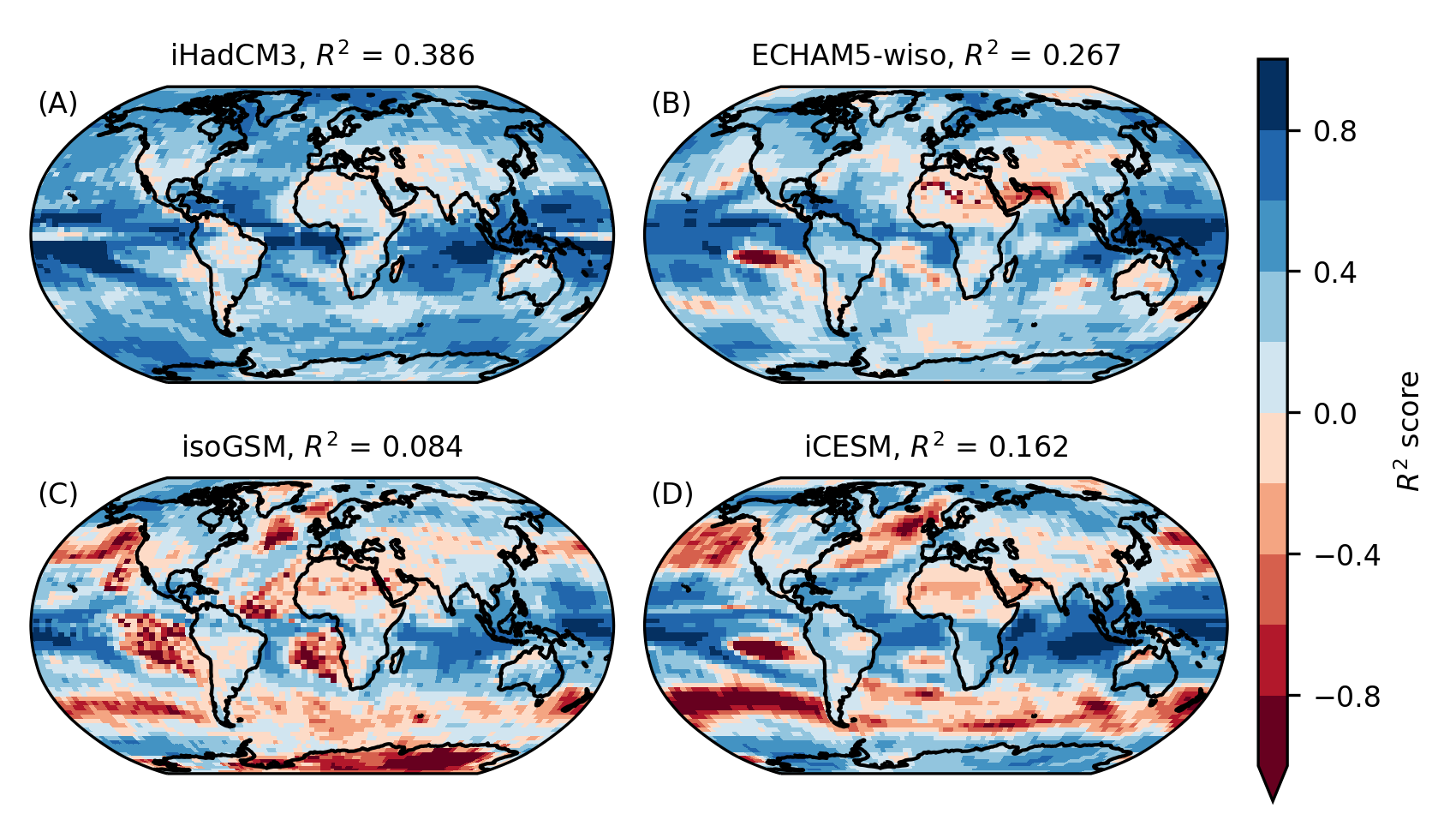}}
{\caption{Results for the cross-prediction task: A UNet is trained on the iHadCM3 training data set. The performance is then evaluated on the test set of various climate models, shown $R^2$ scores are averages over ten runs}
\label{fig:crossprediction}}
\end{figure}

For practical applicability, it is essential that an emulator's performance is robust under varying climatic conditions and under potential biases of the climate model that produces the training data for the emulator. We address these questions by testing how well our emulation generalizes to data generated with different climate models (iCESM, ECHAM5-wiso, isoGSM). To do so, we train the best model architecture so far, the deeper modified flat UNet, on data from iHadCM3. Subsequently, the trained network is used to emulate \dO for the test sets of the other climate model data sets. Results of the emulation are visualized in \Cref{fig:crossprediction}. For all data sets the mean $R^2$ score is positive, meaning that in the global average, the emulation is preferable to predicting the mean state of the corresponding training set. The $R^2$ score is highest for the ECHAM5-wiso simulation and lowest for isoGSM, where $80\%$ less variance is explained than on iHadCM3. 

In all three cross-prediction cases, the performance drops strongly in the Pacific Ocean west of South America, a region that is important for the El Niño–Southern Oscillation (ENSO). This might hint at inter-model differences in the spatial pattern of ENSO variability. For isoGSM, the emulation quality over Antarctica is considerably worse than for all other models. The Antarctic in isoGSM is much less depleted in \dO (less negative \dO) than in the other models while showing similar equator to pole temperature gradients \citep{buhlerInvestigatingStableOxygen2022}. This can potentially impact the relationship between the temporal variations of temperature and \dO.

 For isoGSM and iCESM, $R^2$ is negative over large areas of the mid-latitude oceans. As synoptic-scale variability of moisture transport pathways might be an important factor for \dO in the mid-latitudes, adding predictor variables that encode information on the atmospheric circulation in the respective models could improve the results. The independence of the isoGSM and iCESM runs in these regions must be assessed carefully: isoGSM is forced by sea-surface temperatures and sea-ice distributions of a last millennium run with CCSM4, which is a predecessor model of iCESM. Therefore, characteristics of iCESM might also be present in the isoGSM results. 

We also test, how well the baseline ML models generalize when employed to estimate \dO for other climate models. The very simplistic pixel-wise linear regression yields better results than the PCA-regression baseline. \Cref{fig:crossprediction_linreg} shows the cross-model performance of the linear regression baseline. While the $R^2$ score for iHadCM3 itself is significantly smaller than for the UNet model, the loss of performance when doing cross-prediction is much smaller, for iCESM the results are even better than those obtained with the UNet model. Especially over mid-latitude oceans, the $R^2$ scores of the linear regression are better the ones obtained with the UNet.

\section[Discussion]{Discussion}
\label{sec:discussion}
In a first step towards data-driven emulation of water isotopes in precipitation, we show that in a simulated data set $40\%$ of the interannual \dO variance can be explained by ML models. The emulation quality follows patterns of the correlation between \dO and the predictor variables precipitation amount and surface temperature. This hints at the possibility of further improving the emulation by including other variables that are statistically connected to \dO as predictors. \dO composition depends on atmospheric moisture transport, which in turn, depends on atmospheric circulation. Thus, variables encoding information on atmospheric circulation such as sea-level pressure are promising candidates which should be explored in future research. This could be particularly relevant in the mid-latitudes, where the comparably poor performance of the emulators might be due to synoptic-scale moisture transport variability which is not well captured by annual or monthly means of precipitation and temperature. In addition, relative humidity seems a promising candidate as it is important for the evolution of \dO during the evaporation process.

It should be noted that correlation structures between predictor variables and \dO are likely timescale dependent. Our results suggest that temperature, precipitation and atmospheric circulation variations due to internal variability in the climate system and short-scale external forcing such as volcanic eruptions and solar variability are the most important factors controlling interannual \dO variability. On the other hand, changes in long-term external forcings such as greenhouse gas concentrations and Earth's orbital configuration, and variations in oceanic circulation have been found to explain \dO changes on millennial and orbital (10,000 years and longer) timescales \citep{he_hydroclimate_2021}. This varying importance of factors controlling climate variations can also result in timescale-dependent relationships between the predictor variables surface temperature and precipitation amount \citep{Rehfeld2016}, which limits the generalization of emulators between timescales. Meanwhile, on timescales from hours to weeks, the memory in the atmosphere is higher. Thus, taking into account previous time steps and explicitly tracking moisture pathways for example in tropical or extratropical cyclones could improve the emulation performance. On these timescales, ML methods to model sequences of data, like long short-term memory (LSTM), recurrent neural networks (RNNs), or transformer models could be good alternatives.

A tested spherical CNN architecture shows no clear benefit over a modified version of the standard flat UNet for our task of emulating \dO in precipitation globally. We suppose that this is partly due to the strong latitudinal dependence of the statistical relationships between \dO and the predictor variables (as indicated by correlations in \Cref{fig:iHadCM3_stats}\textsf{C}). Thus, the strength of the spherical network architecture, namely its equivariance to rotations, possibly does not offer a strong benefit. Additionally, the interpolation between the plate carrée grid and the icosahedral grid deteriorates the results. This might be remedied by ``differentiating through'' the interpolation or directly learning the interpolation, as is done in \cite{lamGraphCastLearningSkillful2022}. Using ML architectures that are equivariant to approximate symmetries in the Earth system might still be beneficial in many applications, since adapting the ML approach to symmetries of the problem reduces overfitting and the demands for training data. One might for example use \citet{cohenGroupEquivariantConvolutional2016} as a starting point and test a network that is equivariant under rotations around the polar axis and reflections on the equatorial plane.

The cross-model emulations can be seen as a supplement to test for the generalization to the (unavailable) real-world \dO data. Assuming that each model possesses deficiencies in its \dO simulation, robustness under varying models would hint at robustness in the generalization to real-world data. Additionally, reliable \dO emulations for climate models that do not possess an implementation of water  isotopologues would ideally be done with an emulator that does not overfit to a certain climate model it was trained on. Two reasons that might make an ML emulator perform poorly under cross-emulations are a) weak statistical connections between \dO and the predictor variables in the training set and b) differences in the statistical connections of \dO and the predictor variables between climate models. We investigate whether drops in cross-prediction performance can be attributed to these causes in \Cref{ssec:crossprediction_performance_eval} and \Cref{fig:corrs_diag}. Indeed, most regions, in which there is a drop in emulation performance, coincide with regions of differing correlation structures or weak correlations between \dO and the predictor variables in the iHadCM3 data set (\Cref{fig:corrs_diag}, \textsf{B4} to \textsf{D4} and \textsf{B2} to \textsf{D2}). In regions with weak correlations between \dO and the predictor variables in the iHadCM3 data set such as the Southern Hemisphere mid-latitudes, the UNet has to predict \dO based on spatial similarity structures (teleconnections). The poor performance in the Southern Hemisphere mid-latitudes in IsoGSM and iCESM suggest that the spatial similarity structures differ between those two GCMs and iHadCM3. Here, predictors that encode atmospheric circulation more directly such as sea-level pressure could be beneficial in future studies.

This interpretation is supported by a much sharper drop in performance of the UNet architectures than simple linear regression when methods were trained on the iHadCM3 climate model and then used to emulate other climate model data. As a result, the $R^2$ scores on the other climate models were comparable between UNet and linear regression. This suggests that the UNet might overfit to the spatial anomaly patterns in iHadCM3 given the limited information provided by the predictor variables. This overfitting will partly reproduce deficiencies of the respective data set used for the training of the emulator. It was shown previously that the models used in our study differ in their mean climate state. For example, iHadCM3 and ECHAM5-wiso show a similar global temperature state but iHadCM3 \dO is much more negative in the global mean \citep{buhlerInvestigatingStableOxygen2022}. Similar differences in the spatial anomaly patterns between models need to be explored further to understand their contribution to poor cross-model emulation performance. To obtain a more robust emulator that is applicable across models, one might utilize data from multiple climate models and climate states (e.g. Last Glacial Maximum, mid-Holocene, Pliocene) in the training set.

Spatially, ML estimates are smoother than the true simulated data. The ground truth data show very noisy behavior over dry regions, part of which is likely due to numerical instabilities in the computation of \dO for very low precipitation amounts. Missing data points also occur more frequently in these regions, thus, potentially biasing the emulator and its measured performance. Because of these inconsistencies in the input data, it might be beneficial to focus on particular regions when developing an emulator with the aim of comparing to a certain natural climate archive. Examples are the polar regions for comparisons to ice core data or the mid-latitudes for speleothem records. Restricting the spatial extent would also alleviate artifacts of the map projection and render spherical approaches unnecessary. Alternatively, one might think about the application of ML to do in-painting of missing values of \dO for the training of emulators, similar to \cite{kadowArtificialIntelligenceReconstructs2020}. In this case, the incomplete \dO would serve as an input to the ML method in addition to precipitation and temperature.

Training an isotope emulator on real-world data would avoid uncertainties originating from climate models and the implementation of isotopes within them. Databases of observed \dO in precipitation \citep{IAEAWMO2020} or \dO from natural climate archives \citep{Konecky2020} are publicly available. However, challenges arise from the spatial scarcity and unequal distribution of data, and the short temporal coverage of observations. Here, using graph networks like the one developed by \citet{defferrardDeepSphereGraphbasedSpherical2020} might be an option, and likely strong prior constraints would need to be used to compensate for small data set sizes. For the future goal of comparing emulations to \dO measured in natural climate archives, archive-specific processes need to be taken into account. This is because \dO in precipitation is not archived directly, but always as the response of a sensor of the archiving medium. For example, precipitation \dO is archived in speleothem records as calcite carbonate in accumulating layers that form from cave drip water \citep{Fairchild2012}.

We calculate yearly \dO as the unweighted average of monthly \dO. In most natural climate archives, yearly \dO is weighted by precipitation amount. We tested the influence of such a weighting and find that it does not impact the emulator performance negatively (not shown). However, climate archives can also show seasonal preference in their sensitivity to \dO \citep{Wackerbarth2010, Fohlmeister2017, Baker2019} such that there is likely no optimal way for computing yearly values. Including archive-specific processes could either be a second step in a two-step approach, where an ML emulator is trained to predict \dO in precipitation and then a proxy system model \citep{Evans2013} is used to forward-model archive-specific processes. Alternatively, one might include a differentiable proxy system model in the ML pipeline. This would make it possible to train the ML architecture directly with proxy data instead of \dO measured in precipitation.

\section{Conclusion}
In this study, we explored the ability of machine learning methods to emulate oxygen isotopes as simulated by isotope-enabled general circulation models (GCMs). Focussing on interannual variability in a last millennium simulation, we show that UNet neural networks improve the emulation performance compared to baseline methods such as pixel-wise linear regression and PCA-regression. Averaged over all grid-cells, our best-performing UNet architecture explains $\sim$40\% of the temporal \dO variance. The emulation performs best in polar regions, where \dO is strongly controlled by surface temperature variations, and in low latitude ocean areas, where \dO is highly correlated with precipitation amounts. Lowest performances occur in arid regions, partly because of numerical instabilities in the simulation of \dO for very low precipitation amounts. Using a spherical network architecture does not improve the results compared to a modified flat architecture which accounts better for Earth's spherical geometry than a default UNet architecture. This might be because our spherical UNet architecture is not optimized to capture latitudinal dependences in the relationships between \dO and the predictor variables.

We tested the generalization of the emulator trained on output from the iHadCM3 GCM to last millennium simulations with other GCMs. While the performance is better than predicting the model's climatology for all GCMs, the explained variance is substantially lower than for iHadCM3. Performances are especially poor in regions where the correlation structure between \dO and the predictor variables differs from the correlation structure in iHadCM3 and in regions with low correlations between \dO and the predictor variables in iHadCM3. In the latter case, the UNet architecture learns spatial dependence structures to improve the emulation of \dO. This improves the performance within iHadCM3 compared to pixel-wise regression. However, these spatial structures seem to differ too much between GCMs to facilitate skillful cross-model emulations, especially in the mid-latitudes where encoding synoptic-scale circulation variations could be important to capture \dO variations.

To further improve emulation performance, adding more predictor variables could be a promising next step. In particular, variables such as sea level pressure, which capture characteristics of the atmospheric circulation more directly than surface temperature and precipitation amount, could help in regions with currently poor performance. To compare emulated isotopes to \dO measured in natural climate archives such as ice cores and speleothems, a way of incorporating archive-specific processes needs be investigated. This could be done by incorporating differentiable proxy system models into UNet architectures or by applying proxy system models to the emulator output in a two-step approach. For comparison with \dO measurements in natural climate archives, the timescales of \do variations  recorded by the archives are important. While we focused on interannual timescales in this study, shorter as well as longer timescales could be explored in future research to understand the importance of synoptic-scale processes, local predictor variables and external forcings for \dO emulation across timescales.


\begin{Backmatter}

    \paragraph{Acknowledgments}
    We thank Nadine Theisen for help with the implementation and testing of baseline models.
    We are grateful for the contribution and standardization of model data from Jesper Sjolte, Kei Yoshimura, Madhavan Midhun, Martin Werner, and Josefine Axelsson. 
    
    \paragraph{Funding Statement}
    This research was supported by grants from the Deutsche Forschungsgemeinschaft (DFG, German Research Foundation) through the STACY (project no. 395588486) and CLIMAIC (project no. 442926051) projects.

    \paragraph{Competing Interests}
    None.
    
    \paragraph{Data Availability Statement}
    The data used in this study can be freely downloaded here: \url{https://doi.org/10.5281/zenodo.7516327}. Code to reproduce our experiments is publicly available at \url{https://github.com/jonathanwider/isoEm}, it is subject to the license statements in the GitHub repository. 
    
    \paragraph{Ethical Standards}
    The research meets all ethical guidelines, including adherence to the legal requirements of the study country.
    
    \paragraph{Author Contributions}
    Conceptualization:  UK, KR, JW, NW; Data curation: JB, KR; Formal Analysis: JW; Funding Acquisition: KR; Investigation: JW;  Methodology: JK, UK, KR, JW, NW; Project administration: UK, KR; Resources: KR; Software: JW; Supervision: JB, JK, UK, KR, NW; Validation: JK, JW; Visualization: JW; Writing original draft: JK, JW; Writing – review \& editing: all authors. All authors approved the final submitted draft.
    
    \paragraph{Supplementary Material}
    Supplementary material has only been provided as an Appendix to this document.

    \bibliographystyle{apalike}
\bibliography{lit.bib}

\end{Backmatter}

\clearpage
\begin{appendix}\appheader
\section{Supplementary Material}\label{appendixA}

\renewcommand{\thefigure}{A.\arabic{figure}}
\setcounter{figure}{0}
\renewcommand{\thetable}{A.\arabic{table}}
\setcounter{table}{0}

\subsection{Implementation Details}
\label{ssec:implementation_details}
We use pytorch \citep{paszkePytorchImperativeStyle2019} to implement the CNNs and scikit-learn \citep{pedregosaScikitlearnMachineLearning2011} for the baseline models. Our code is publically available on GitHub.\footnote{\url{https://github.com/jonathanwider/isoEm}.}

\subsubsection{Validation Experiment}
\label{sssec:validation}
\begin{figure}[t]%
    \FIG{\includegraphics[width=\textwidth]{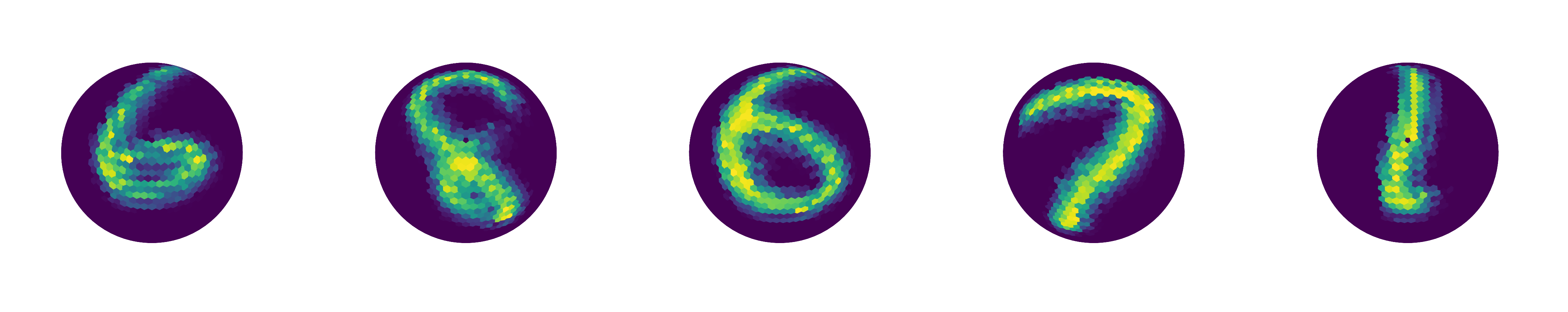}}
    {\caption{Validation task for the icosahedral neural network: shown are randomly selected examples of handwritten digits from the MNIST data set, projected onto the icosahedral grid, in this case the refinement level of the icosahedral grid $r$ equals 4. For the validation tasks, three versions of this data set are generated: One where the digits are not rotated ("N"), one in which rotations of the symmetry group of the icosahedron are applied ("I") and one in which we apply rotations of the group of rotations of the sphere("R")}
    \label{fig:MNIST_digits}}
\end{figure}

The validation experiment is a variation of one of the seminal tasks in ML: the classification of the MNIST data set of handwritten digits \citep{lecunMnistDatabaseHandwritten2005}. As a toy example, \citet{cohenGaugeEquivariantConvolutional2019} project the digits onto the southern hemisphere of the icosahedral grid, some examples are visualized in \Cref{fig:MNIST_digits}. Then, three types of data sets are produced: the digits can either be left as they are (``non-rotated'') - or the icosahedral grid can be rotated before the digits are projected onto it, either by a symmetry transformation of the icosahedron or by an unconstrained rotation (``fully-rotated''). Because the network architecture of \citet{cohenGaugeEquivariantConvolutional2019} is equivariant to rotations of the icosahedron, the network should be able to classify digits that have been rotated by elements of this symmetry group with the same accuracy as non-rotated digits, even if during training it was never shown any rotated digits. 

\begin{figure}[t]%
\FIG{\includegraphics[width=\textwidth]{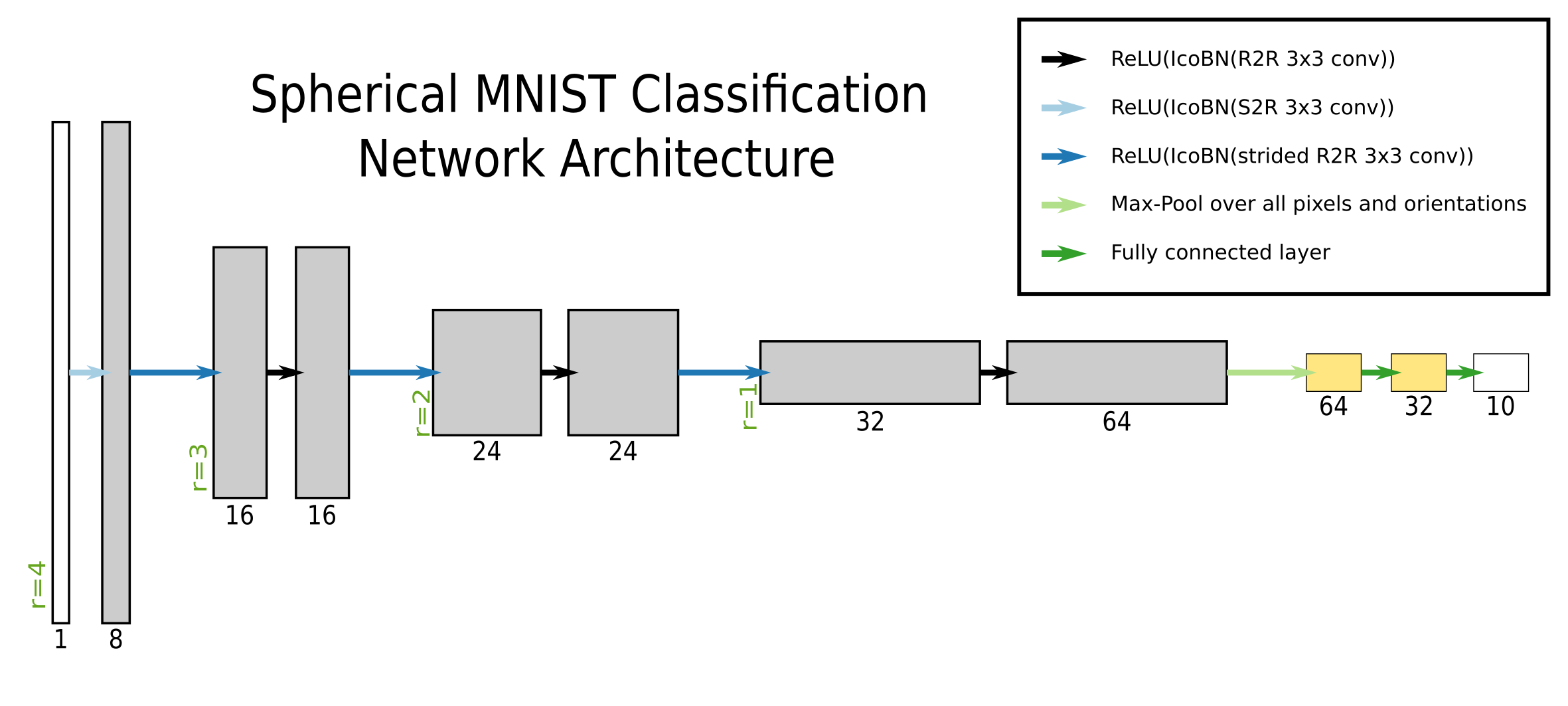}}
{\caption{Sketch of the architecture used to validate our implementation of the icosahedral network of \citet{cohenGaugeEquivariantConvolutional2019}. Spatial resolutions (green text) relate to the refinement level of the icosahedral grid. The architecture was chosen according to the details given in the supplementary material of the original paper. "S2R" and "R2R" are specific types of convolutional layers developed to maintain equivariance, and IcoBN is an adapted version of batch normalization. For details, see \citet{cohenGaugeEquivariantConvolutional2019}}
\label{fig:validation_architecture}}
\end{figure}

The data set creation process, our network architecture (\Cref{fig:validation_architecture}), and the training procedure for this task follow \citet{cohenGaugeEquivariantConvolutional2019} as closely as possible. We train for 60 epochs on the non-rotated and fully-rotated data sets and for one epoch on the icosahedrally rotated data set, which is 60 times larger than the other sets. We use a cross-entropy loss function and the Adam optimizer \citep{kingmaAdamMethodStochastic2014} with a learning rate of $0.001$ and the other parameters at their default values $\beta_{1,2}=(0.9, 0.999), \epsilon=10^{-8}$.

Our results in \Cref{tab:validation_experiment} are comparable or better than those reported in \cite{cohenGaugeEquivariantConvolutional2019}. Notably, they demonstrate the equivariance of our implementation - the classification results for digits rotated by an icosahedral symmetry are almost identical to those of non-rotated digits.\footnote{We did not fix a random seed, therefore, the results of N/N and N/I are not exactly identical.} Small differences between the results of our implementation and the implementation of \cite{cohenGaugeEquivariantConvolutional2019} can partly be attributed to the fact that none of our hyperparameters were tuned - and to implementation details not described in \citet{cohenGaugeEquivariantConvolutional2019}, like the batch size or the used optimizer.

\begin{table}[t]
\tabcolsep=0pt%
\TBL{\caption{Classification accuracies (fraction of correctly classified to the total number of tested digits) for the icosahedral MNIST validation experiment \citep{cohenGaugeEquivariantConvolutional2019}. "N", "I" and "R" indicate data sets, where either no rotations, rotations of the symmetry group of the icosahedron, or rotations of the symmetry group of the sphere were applied. Shown results are averages over three runs.}
\label{tab:validation_experiment}
}
{\begin{fntable}
\begin{tabular*}{\textwidth}{@{\extracolsep{\fill}}lcccccc@{}}\toprule%
\TCH{Training set rot. type/Test set rot. type} & \TCH{N/N} & \TCH{N/I} & \TCH{N/R} & \TCH{I/I} & \TCH{I/R} & \TCH{R/R}\\\midrule
\TCH{Cohen et al.} &\textbf{99.43}&\textbf{99.43}&\textbf{69.99}&\textbf{99.38}&66.26&99.31\\ 
\TCH{Ours} &99.23&99.34&68.57&99.27&\textbf{69.31}&\textbf{99.37}\\
\botrule
\end{tabular*}%
\end{fntable}}
\end{table}

\subsubsection{Isotope Emulation Experiment}
\label{sssec:emulation}

\paragraph{UNet Training}
To optimize the UNet models, we use the Adam optimizer \citep{kingmaAdamMethodStochastic2014}, with $\texttt{lr} = 0.001$ and the other parameters at their default values: $\beta_{1,2}=(0.9, 0.999), \epsilon=10^{-8}$. We use early stopping with a patience of 5, i.e. we abort training if no global minimum of the validation set loss is reached in 5 consecutive epochs. As a result, the training runs are stopped after roughly 20 epochs. We use a batch size of 8. Training a model takes on the order of minutes to tens of minutes for a single run on a basic graphics card (NVIDIA GeForce GTX1650).

For both icosahedral and flat UNets, we use batch normalization \citep[BN, ][]{ioffeBatchNormalizationAccelerating2015} and ReLU-activations after all but the last convolutional layer. We use 2x2 max-pooling to go to coarser spatial resolution and nearest neighbor interpolation (plate carrée projection) and a modified form of bilinear interpolation (icosahedral grid) to increase resolution. Data from skip connections and upsampling are simply concatenated.

\paragraph{Icosahedral UNet Approach}
We create the icosahedral data set using an icosahedral grid at refinement level $r=5$, meaning we apply 5 recursive steps in which each of the triangular faces of the icosahedron is divided into four smaller triangles, with the grid points subsequently projected back to the spherical surface. This results in a grid with $5 \times 2^{2r+1}+2 = 10242$ vertices, while the iHadCM3 latitude-longitude grid encompasses 6816 grid points. Because the areas covered by the iHadCM3 grid cells vary with latitude, the resolution of the icosahedral grid is higher than the resolution of the iHadCM3 grid close to the equators, while close to the poles iHadCM3 data is better resolved than the icosahedral grid.

\begin{figure}[t]%
\FIG{\includegraphics[width=\textwidth]{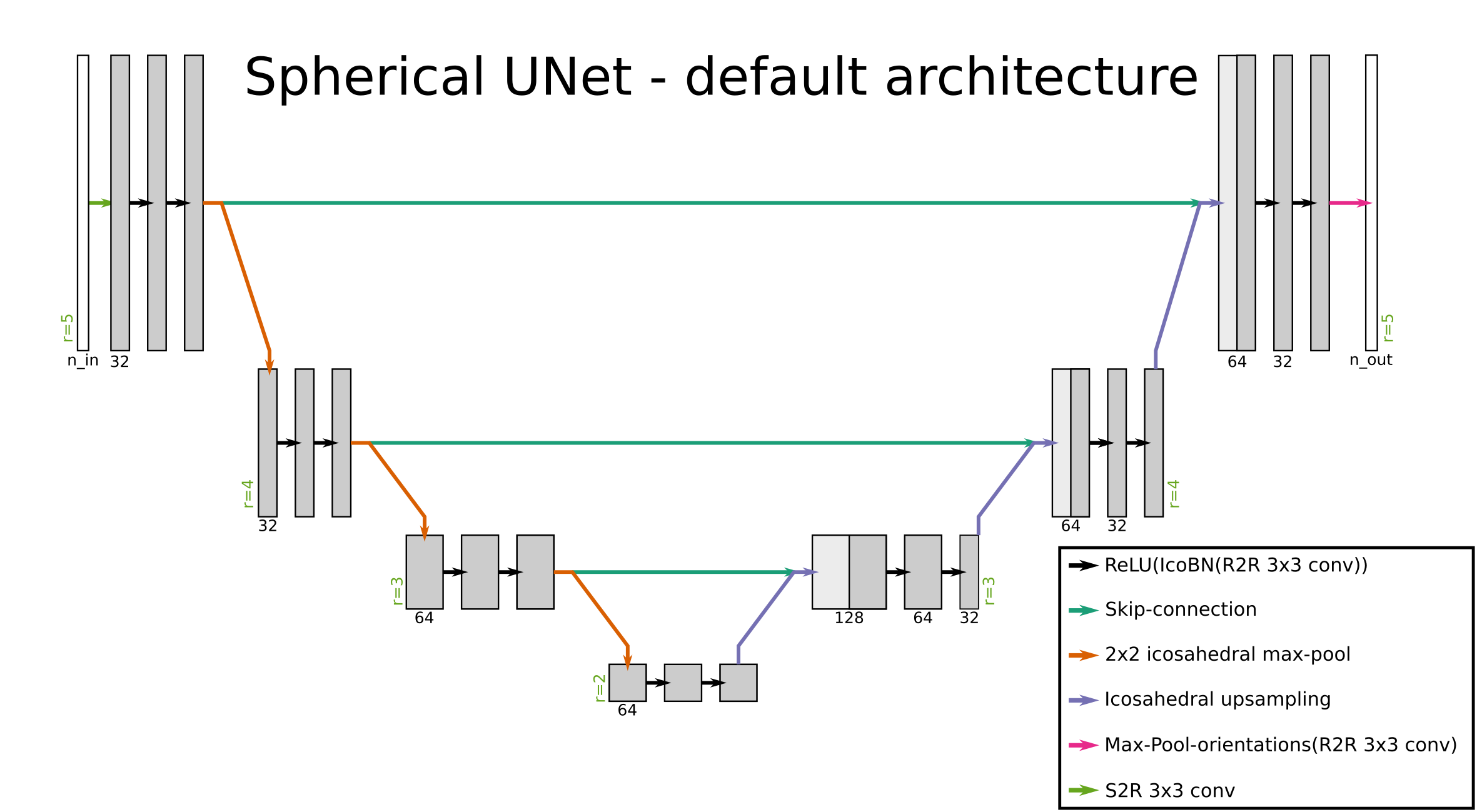}}
{\caption{Sketch of the default icosahedral UNet architecture that takes into account the spherical nature of our data. The parameters are chosen similar to details given in the paper of \cite{cohenGaugeEquivariantConvolutional2019}. The icosahedral refinement level $r$ is given in green, the number of features per layer in black below the corresponding blocks. ``S2R'' (scalar-to-regular) and ``R2R'' (regular-to-regular) are convolutional layers defined by \cite{cohenGaugeEquivariantConvolutional2019} to achieve equivariance to a group of symmetry transformations}
\label{fig:unet_ico}}
\end{figure}

The architecture used for the icosahedral grid is visualized in \Cref{fig:unet_ico}. It is inspired by an architecture in \citet{cohenGaugeEquivariantConvolutional2019}. There is an implementation uncertainty remaining that relates to the treatment of the vertices of the icosahedron at $r=0$, i.e. the 12 corners of the unrefined grid. In opposition to all the other pixels in the icosahedral grid, these possess only five neighboring pixels (instead of six), thus introducing irregularities in the convolution patterns. Unable to extract the exact treatment of these corner pixels in \citet{cohenGaugeEquivariantConvolutional2019}, we made the choice to set these corner pixel values to zero in every layer. This will likely introduce biases at very coarse resolutions (i.e. $r$ small).

We use the MSE-Loss from \Cref{eq:weighted_masked_mseloss} without weighting, because all pixels in the icosahedral grid cover an approximately equal area. Padding is done similarly to \citet{cohenGaugeEquivariantConvolutional2019} in a way that asserts continuity between the faces of the icosahedral grid.

\begin{figure}[t]%
\FIG{\includegraphics[width=\textwidth]{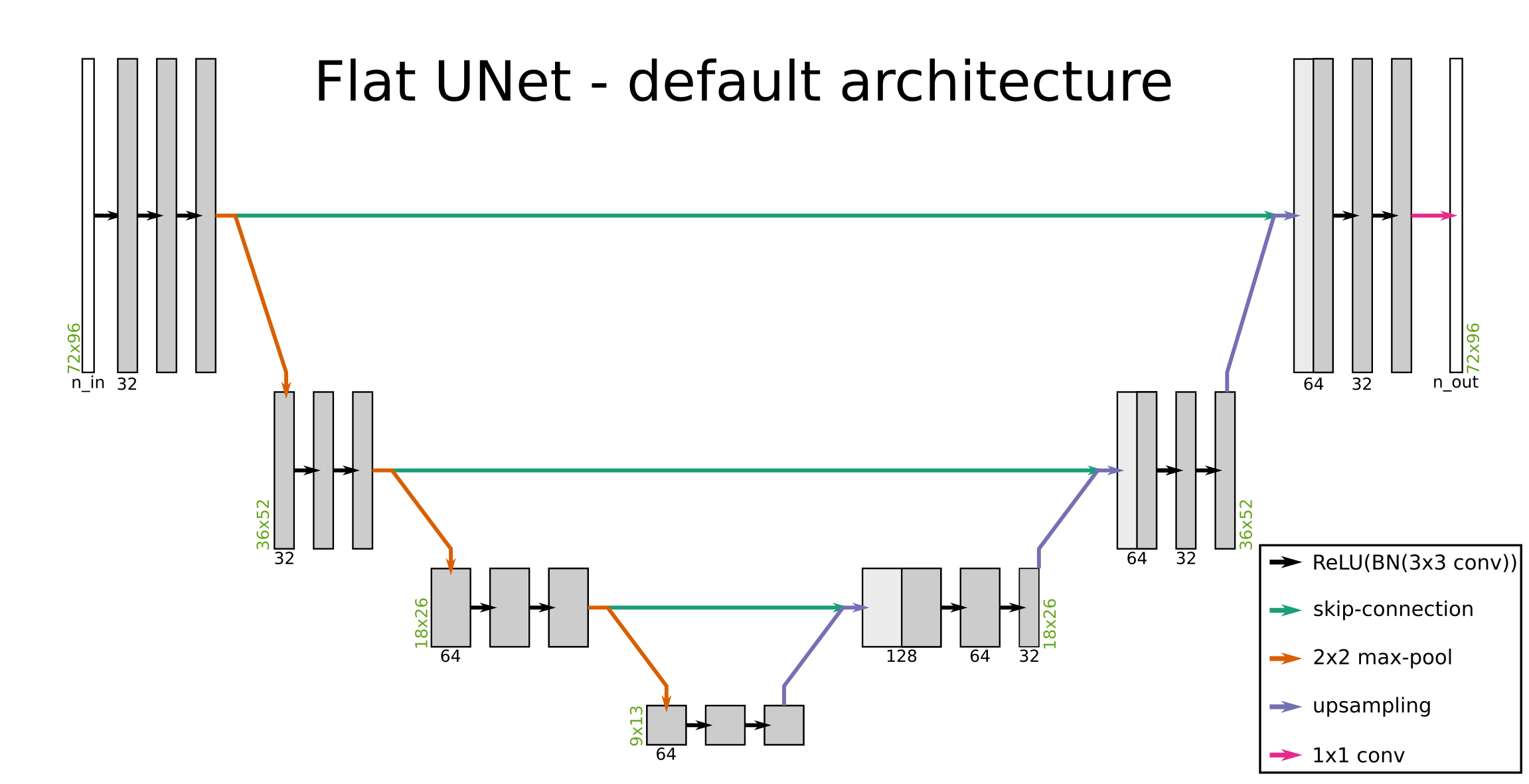}}
{\caption{Sketch of our default flat UNet architecture. The parameters were chosen to roughly match in size with the icosahedral UNet architecture. Before the data is input into the network, it is resized in order to assure divisibility during down-and up-scaling, thus the input resolution (green) to the first layer is $72\times96$  instead of $71\times96$ as in the iHadCM3 grid. In the end, the output of the architecture is scaled back to the original grid. The number of computed features per layer is written below the corresponding data blocks (black). The number of input features ($n_{in}$) depends on the number of chosen variables (it equals 2 if using both temperature and precipitation amount), and the number of output features $n_{out}$ equals $1$ for all our applications. This graphic doesn't include the coordinate features that get appended to the data-tensors before each convolution, if CoordConv \citep{liuIntriguingFailingConvolutional2018} is used. \textup{BN} is short for batch normalization, and \textup{3x3 conv} indicates a convolutional layer with a $3\times3$ convolutional kernel}
\label{fig:unet_flat}}
\end{figure}

\paragraph{Flat UNet}
The default architecture for the flat UNets is visualized in \Cref{fig:unet_flat}. By default, we use zero padding to keep the resolution before and after convolutions identical, in most experiments, however, only the latitudes get zero-padded, while we pad the longitudes cyclically to avoid discontinuities.

\subsection{Investigating reasons for cross-prediction performance drops}
\label{ssec:crossprediction_performance_eval}
\begin{figure}[t]%
\FIG{\includegraphics[width=\textwidth]{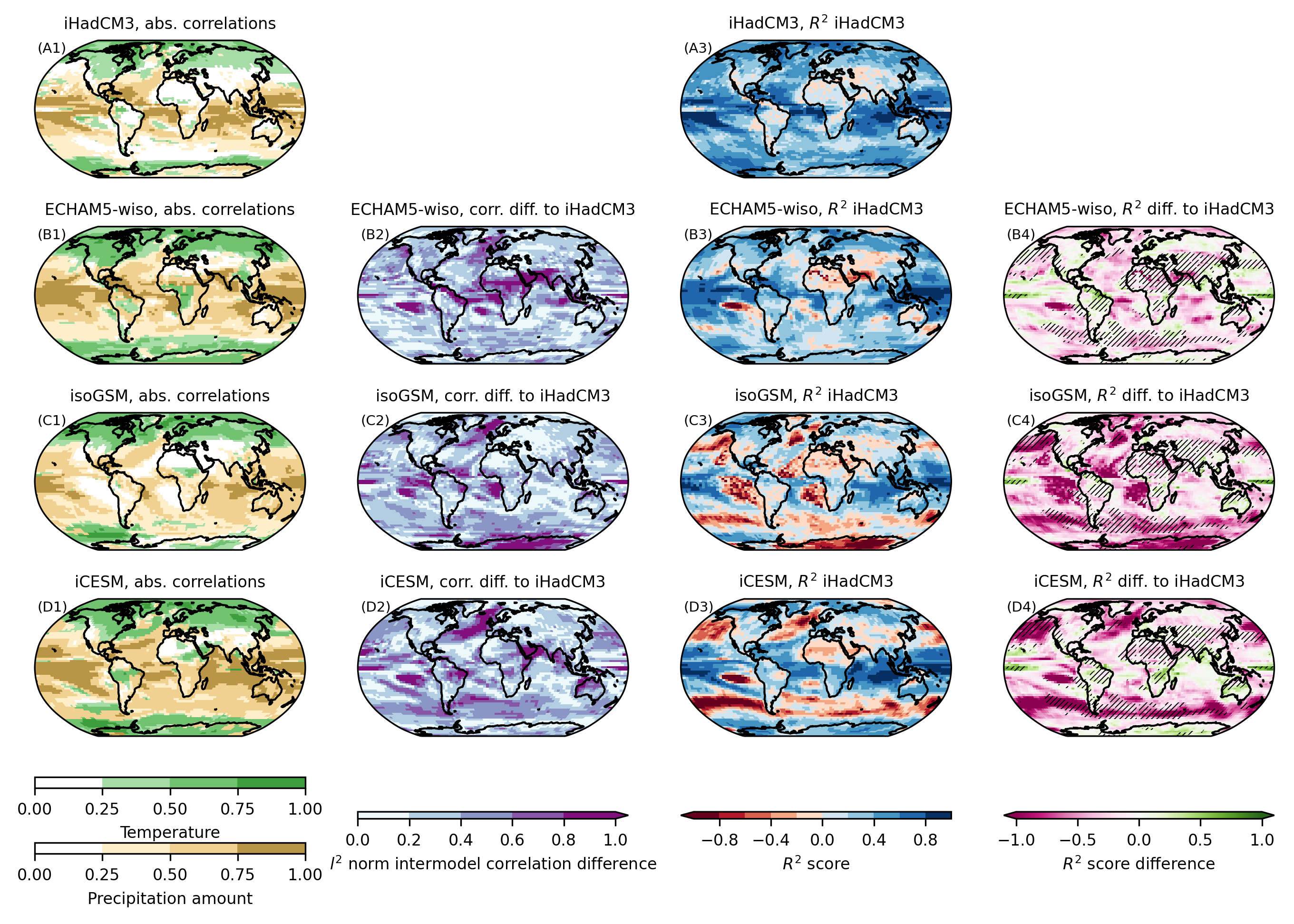}}
{\caption{Investigating reasons for emulation quality decrease in cross-prediction experiments: \textsf{\textup{(A1)}} to \textsf{\textup{(D1)}} show the absolute correlations between \dO and the predictor variables. For each grid cell, only the larger of the two absolute correlation coefficients is shown. \textsf{\textup{(B2)}} to \textsf{\textup{(D2)}} show the differences in the correlation structure between the models as computed in \Cref{eq:corr_diff}.  \textsf{\textup{(A3)}} to \textsf{\textup{(D3)}} show the cross-prediction $R^2$ scores of the best deeper flat UNet model and are identical to the content of \Cref{fig:crossprediction}. \textsf{\textup{(B4)}} to \textsf{\textup{(D4)}} show $R^2$ scores differences between each model and iHadCM3. In hatched regions no correlation coefficient has an absolute value bigger than 0.25 in the iHadCM3 data set}
\label{fig:corrs_diag}}
\end{figure}

We investigate the drops in $R^2$ score that occur when predicting with a model trained on iHadCM3 data on simulation data from other climate models. This experiment and its results were described in \Cref{sec:results-cross-comparison}. As described in the main text, a possible explanation might be differences in the statistical connections between \dO and the predictor variables amongst the different climate models. To estimate these differences, we calculate the root of the squared differences in correlation coefficients between each model and iHadCM3 grid-box wise:
\begin{equation}
    \label{eq:corr_diff}
    \left(\left(r_\text{d18O,prec}^\text{Model} - r_\text{d18O,prec}^\text{iHadCM3}\right)^2 + \left(r_\text{d18O,tsurf}^\text{Model} - r_\text{d18O,tsurf}^\text{iHadCM3}\right)^2 \right)^{\frac{1}{2}}
\end{equation}
Here $r$ indicates the (temporal) Pearson correlation coefficients computed for each model and grid box. The results are visualized in panels \textsf{\textup{(B2)}} to \textsf{\textup{(D2)}} of \Cref{fig:corrs_diag}.

Additionally, the generalization to other models might be impaired in regions where there are weak or no statistical relationships between \dO and the predictor variables in the iHadCM3 data. This may occur due to iHadCM3 misrepresenting the \dO relationships in the Earth system - or there might just not be strong connections between \dO and the predictor variables in these regions. In a simplistic approach, we assess this by plotting regions in which neither the correlation of temperature to \dO nor the correlation of precipitation amount and \dO exceeds an absolute value of $0.25$ as hatches in \Cref{fig:corrs_diag} \textsf{\textup{(B4)}} to \textsf{\textup{(D4)}}.

In panels \textsf{\textup{(B4)}} to \textsf{\textup{(D4)}}, we plot the difference in cross-prediction performance between each model and iHadCM3, whose training set was used to train the emulator. We observe that especially for the ENSO region west of South America and over the North Atlantic Ocean 
\citep[a region relevant for the North Atlantic Oscillation, see ][]{wannerNorthAtlanticOscillation2001} a decline in performance coincides with changes in the correlation structure between the models. Over the southern oceans, there are both changes in correlation structure between the models and weak correlations for iHadCM3.

\begin{figure}[t]%
\FIG{\includegraphics[width=\textwidth]{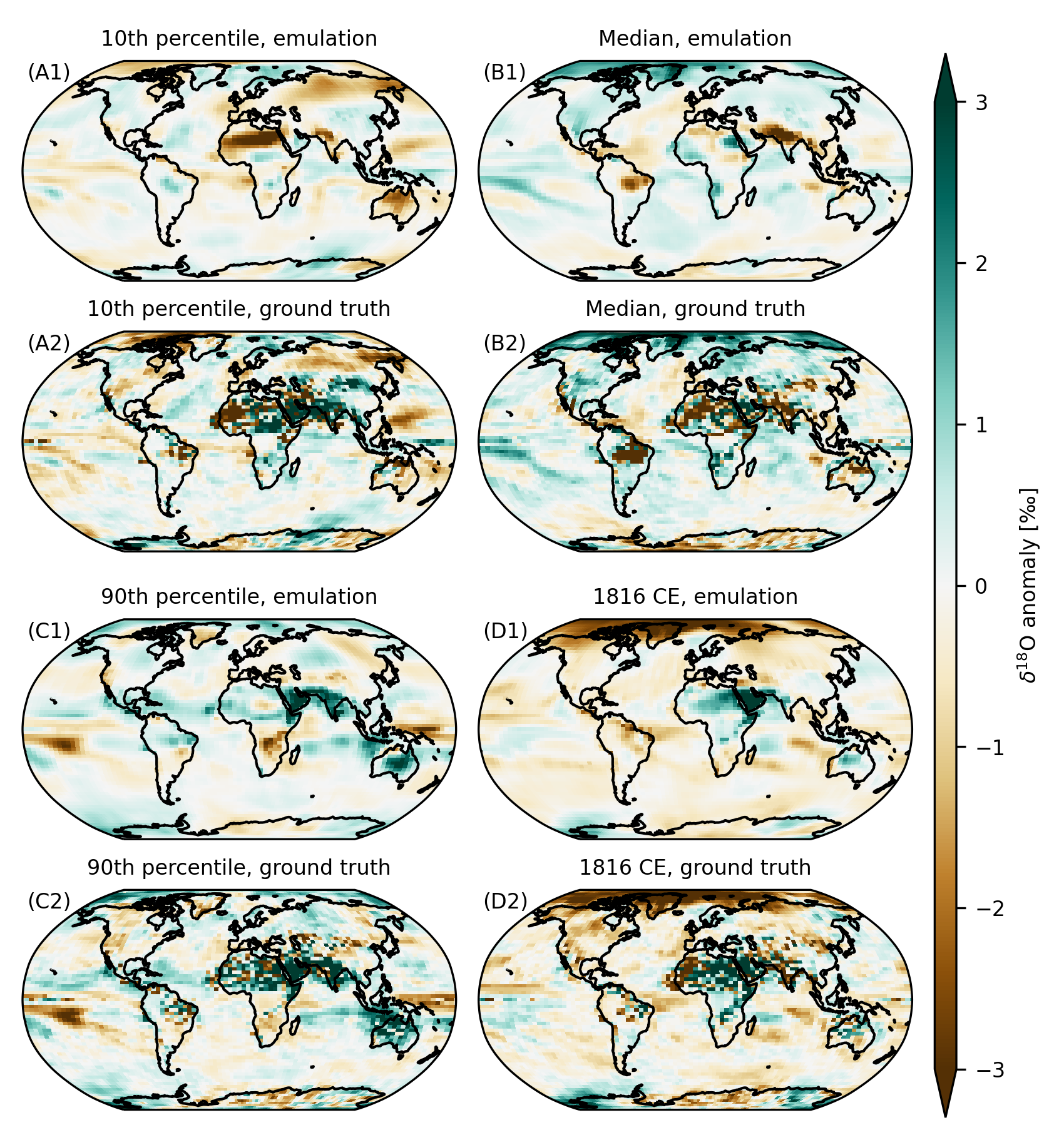}}
{\caption{Emulation results on iHadCM3 data set: We show anomalies produced by the ML-emulator (``emulation'') and the ``true'' result in the iHadCM3 data set (``ground truth''). The anomalies are computed as the difference to the training set mean. We select time steps in which the emulator performs especially strong (``90th percentile'') and weak (``10th percentile''), as measured by the anomaly correlation coefficient (ACC). Additionally, we show the time step, for which the ACC reaches its median value, and a year with a pronounced climatic anomaly: 1816 CE, the ``year without a summer''}
\label{fig:iHadCM3_anomalies_all}}
\end{figure}

\begin{figure}[t]%
\FIG{\includegraphics[width=\textwidth]{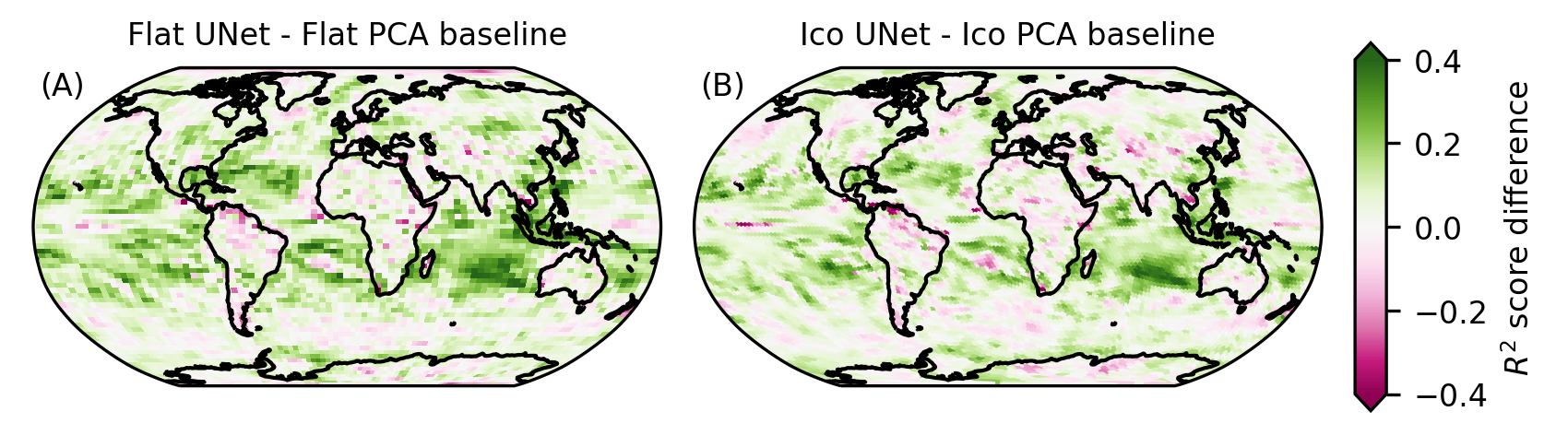}}
{\caption{The difference in $R^2$ score between the UNet model and the PCA-based regression baseline for each grid type,\textsf{\textup{(A)}}: plate carrée grid, \textsf{\textup{(B)}}: icosahedral grid. On the plate carrée grid, we use the ``modified'' version of the flat UNet, including the modifications remedy distortions due to the spherical nature of the data. Green colors indicate better performance of the UNet compared to the baseline model. Before computing the differences, $R^2$ scores of ten runs are averaged for each configuration}
\label{fig:differences_to_baselines}}
\end{figure}

\begin{figure}[t]%
\FIG{\includegraphics[width=\textwidth]{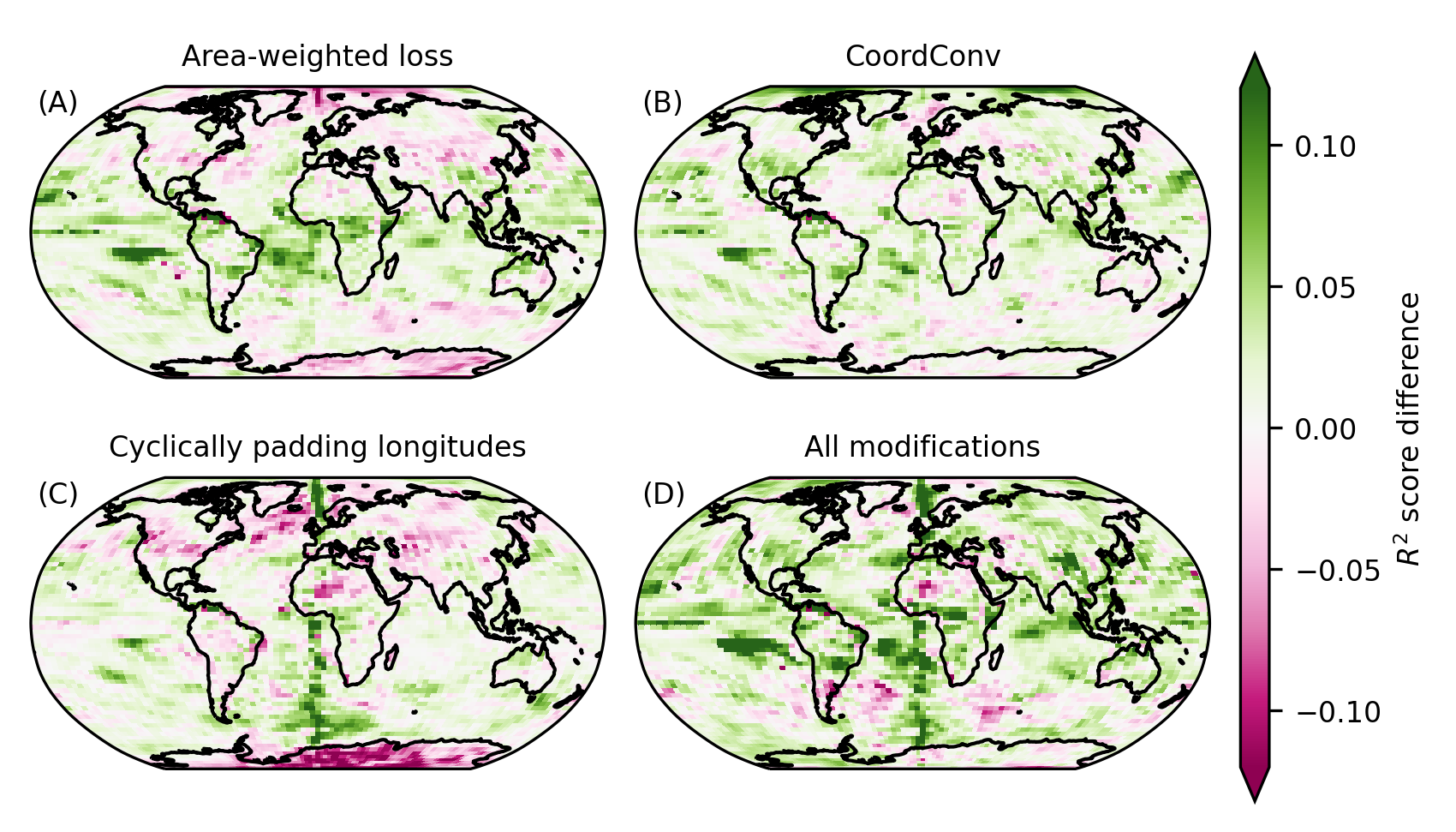}}
{\caption{Effects of modifications to the ``standard'' UNet, which treats the longitude-latitude grid as a flat image: In each panel, we show the difference in $R^2$ score with respect to the unmodified version. We investigate the following modifications: 
\textsf{\textup{(A)}} A loss function, in which the contribution of each grid box is weighted by the area it covers on Earth's surface. \textsf{\textup{(B)}} CoordConv \citep{liuIntriguingFailingConvolutional2018}, a modification to CNNs, that allows the network to access the coordinates of locations in the image and break translational equivariance. 
\textsf{\textup{(C)}} padding the longitudes cyclically instead of using zero padding and 
\textsf{\textup{(D)}} the joint effect of the modifications. Shown in each panel are differences between averages over the $R^2$ scores of ten runs}
\label{fig:modifications_flat_unet}}
\end{figure}

\begin{figure}[t]%
\FIG{\includegraphics[width=\textwidth]{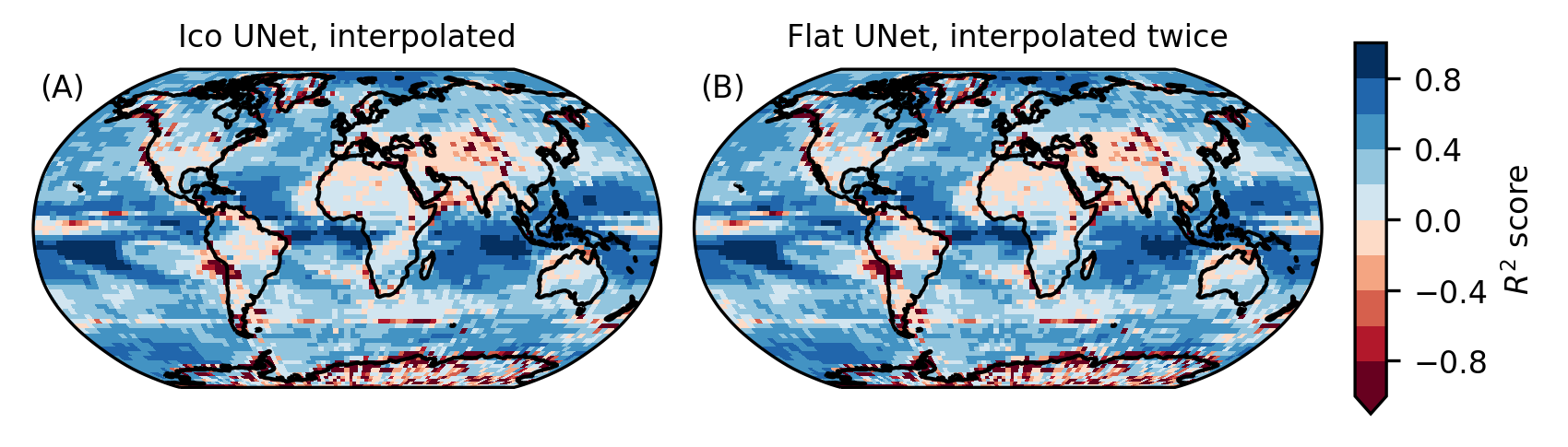}}
{\caption{Interpolation between grids degrades results. \textsf{\textup{(A)}} Performance of the icosahedral UNet architecture, here the results were interpolated to the flat grid. \textsf{\textup{(B)}} results of the flat UNet architecture, after interpolating the predictions to the icosahedral grid and back to the plate carrée grid. The prediction quality is almost indistinguishable from the results of the icosahedral UNet. Additionally, the results are bad in regions, where we expect interpolations to have a negative effect: Over coastal and in the polar regions, where the icosahedral grid cells are larger then the iHadCM3 grid cells and therefore data partly gets ``averaged'' when interpolating from the finer iHadCM3 grid and to the coarser icosahedral grid and back. Thus, we can attribute a large part of the $R^2$ score differences to the interpolations between the grids}
\label{fig:results_flat_grid_interpolation}}
\end{figure}

\begin{figure}[t]%
\FIG{\includegraphics[width=\textwidth]{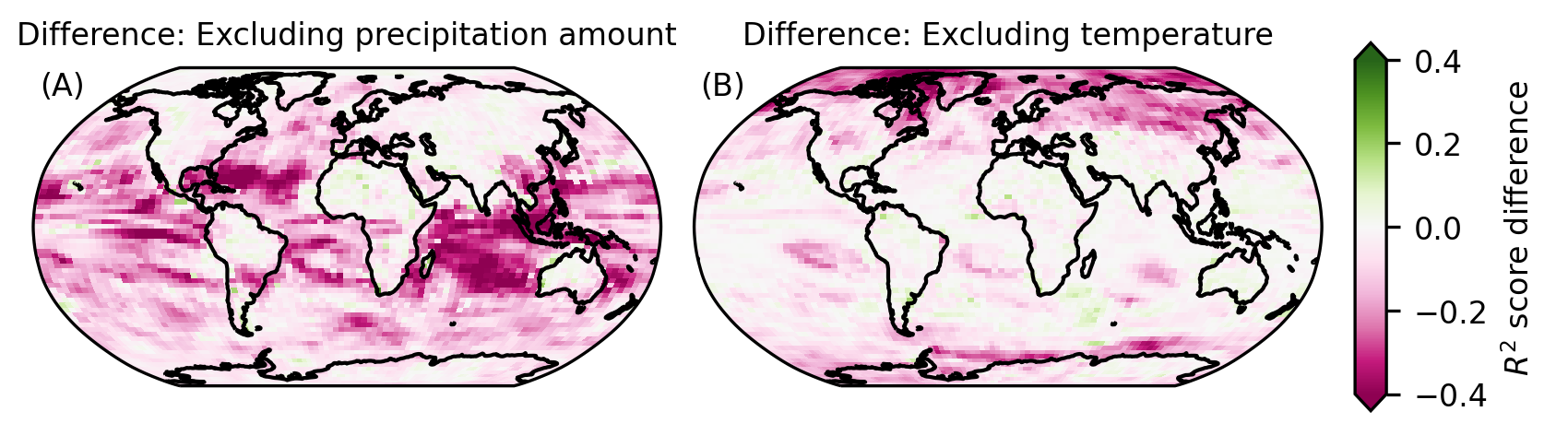}}
{\caption{Difference in emulation quality ($R^2$ score) between using only one of the predictor variables (surface temperature and precipitation amount) and using both simultaneously. Regions shaded in purple indicate a drop in performance when leaving out the corresponding predictor variable. Before computing the differences, averages over the $R^2$ scores of ten runs were formed for each configuration.}
\label{fig:differences_variables}}
\end{figure}

\begin{figure}[t]%
\FIG{\includegraphics[width=\textwidth]{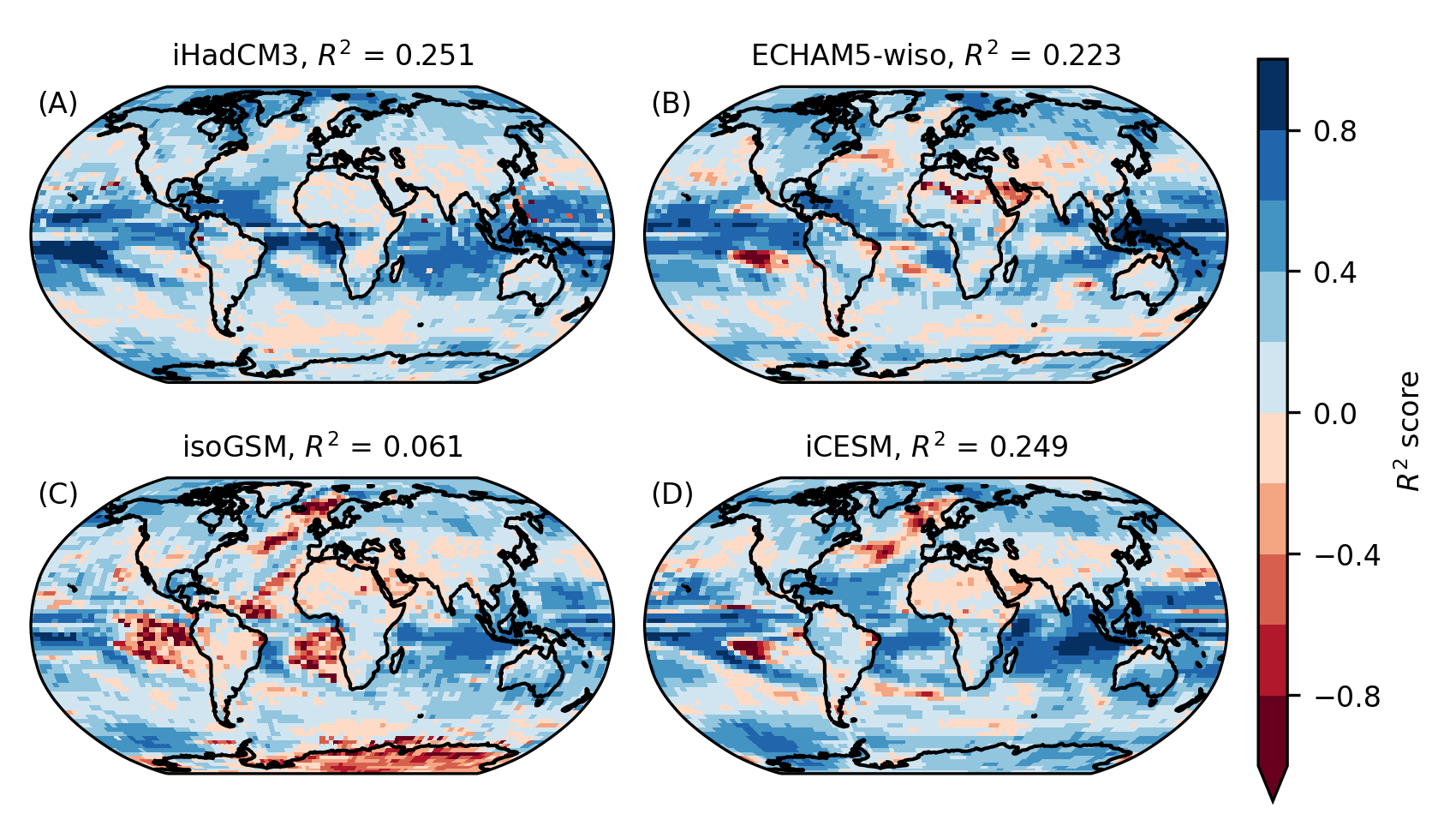}}
{\caption{Results for the cross-prediction task with a pixel-wise linear regression model. The model was trained on the iHadCM3 data set, where the performance is worse than with the UNet model. For cross-predictions on other data sets, however, the emulation quality of the linear model is comparable to or even better than the performance of the UNet, as can be seen by comparing the globally averaged $R^2$scores in panels \textsf{\textup{(B)}} to \textsf{\textup{(D)}} to the corresponding panels of \Cref{fig:crossprediction}}
\label{fig:crossprediction_linreg}}
\end{figure}

\begin{figure}[t]%
\FIG{\includegraphics[width=0.6\textwidth]{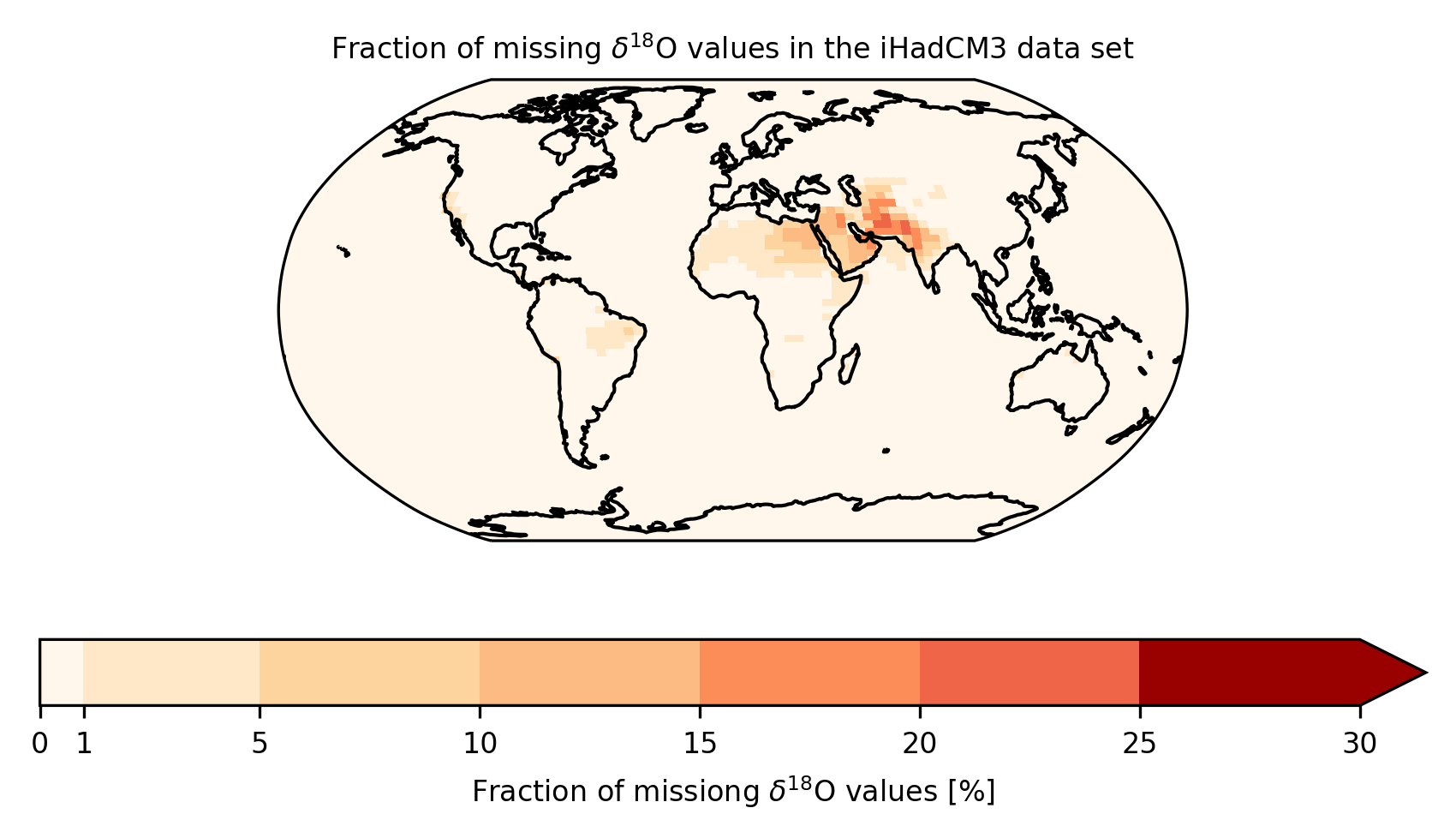}}
{\caption{Investigating missing \dO values: Spatial distribution of missing \dO data for iHadCM3 on monthly timescale. While for over 80\% of the gridboxes, not a single value is missing, missing values are clustering mainly in hot and dry regions. For the worst grid box, \dO is missing in 25\% of the time steps}
\label{fig:missing_pixels}}
\end{figure}

\begin{table}[t]
\tabcolsep=0pt%
\TBL{\caption{Effect of modifications to flat UNet architecure, globally averaged $R^2$ scores. Shown are standard deviations and mean over ten runs.}
\label{tab:modifications}
}
{\begin{fntable}
\begin{tabular*}{\textwidth}{@{\extracolsep{\fill}}cccc@{}}\toprule%
\TCH{Use CoordConv} & \TCH{Use area-weighted loss} & \TCH{Use cylindrical padding} & \TCH{$R^2$score}\\\midrule
No&No&No& $0.352 \pm 0.015$\\
No&No&Yes& $0.357 \pm 0.015$\\
No&Yes&No& $0.365 \pm 0.005$\\
Yes&No&No& $0.367 \pm 0.010$\\
Yes&Yes&Yes& $0.377 \pm 0.005$\\\botrule
\end{tabular*}%
\end{fntable}}
\end{table}

\end{appendix}

\end{document}